\documentclass[11pt]{article}
\usepackage{fullpage}
\usepackage{amsmath,amssymb}
\usepackage{enumitem}
\usepackage{algorithm}
\usepackage[noend]{algpseudocode}
\usepackage{color}
\usepackage{setspace}
\usepackage{wrapfig}
\usepackage{graphicx}
\usepackage{multicol}
\usepackage{chngcntr}
\usepackage{apptools}
\AtAppendix{\counterwithin{lemma}{section}}
\newenvironment{sketch-proof}{%
  \proof}{\hfill{\small$\Box$}\endproof}
  
\algrenewcommand\alglinenumber[1]{\tiny #1:}

\newtheorem{theorem}{Theorem}
\newtheorem{lemma}[theorem]{Lemma}

\newcommand{\prf}[1]{{}}

\newenvironment{Proof}{\noindent{\bf Proof.}}{\hfill$\Box$\FF}

\newtheorem{Def}{Definition}[section]

\newcommand{\sacode}[5]
{ \vspace{.06in} \hrule \vspace{.06in} 
 \noindent {\bf #1}: \\
 \footnotesize \noindent {\bf Signature:}\B \nobreak
 \normalsize \begin{quote} \nobreak #2 \end{quote}
 \footnotesize \noindent {\bf States:}\B \nobreak
 \begin{quote} \nobreak #3 \end{quote}
 \noindent {\bf Transitions:} \nobreak
 \vspace{-.2in} \nobreak
 \normalsize #4
 \vspace{-.06in} \hrule \vspace{.06in} 
}

\newcommand{\act}[1]{%
    \relax\ifmmode
        \mathord{\mathcode`\-="702D\sf #1\mathcode`\-="2200}%
    \else
        $\mathord{\mathcode`\-="702D\sf #1\mathcode`\-="2200}$%
    \fi
}

\newcommand{\tup}[1]{%
    \relax\ifmmode
      \langle #1 \rangle%
    \else
        $\langle$#1$\rangle$%
    \fi
}

\newcommand{\seq}[1]{%
    \relax\ifmmode
      \langle \! \langle #1 \rangle \! \rangle%
    \else
        $\langle \! \langle$ #1 $\rangle \! \rangle$%
    \fi
}

\newcommand{\B}{\vspace*{-\smallskipamount}}

\newcommand{\FF}{\vspace*{\medskipamount}}

\newcommand{\ms}[1]{%
    \relax\ifmmode
        \mathord{\mathcode`\-="702D\it #1\mathcode`\-="2200}%
    \else
        {\it #1}%
    \fi
}

\newcommand{\lit}[1]{%
    \relax\ifmmode
        \mathord{\mathcode`\-="702D\sf #1\mathcode`\-="2200}%
    \else
        {\it #1}%
    \fi
}

\newcommand{\XDK}[1]{}
\newcommand{\remove}[1]{} 
\newcommand{\uselater}[1]{} 

\providecommand{\tup}[1]{%
    \relax\ifmmode
      \langle #1 \rangle%
    \else
        $\langle$#1$\rangle$%
    \fi
}

\renewcommand{\act}[1]{%
    \relax\ifmmode
        \mathord{\mathcode`\-="702D\sf #1\mathcode`\-="2200}%
    \else
        $\mathord{\mathcode`\-="702D\sf #1\mathcode`\-="2200}$%
    \fi
}

\makeatletter
\def\mainlistofsymbols{
  \normalsize
  \vspace*{1.5 em}
  \@starttoc{los}
}

\def\partonelistofsymbols{
  \normalsize
  \vspace*{1.5 em}
  \@starttoc{p1los}
}

\def\parttwolistofsymbols{
  \normalsize
  \vspace*{1.5 em}
  \@starttoc{p2los}
}

\def\l@symbol#1#2{\addpenalty{-\@highpenalty} \vskip 4pt plus 2pt
{\@dottedtocline{0}{0em}{8em}{#1}{#2}}}
\makeatother

\newcommand{\newhiddensym}[2]{%
}

\newcommand{\stateSet}[1]{states(#1)}

\newcommand{\actionSet}[1]{actions(#1)}

\newcommand{\algIOA}[2]{\ifmmode{\text{#1}_{#2}}\else{$\text{#1}_{#2}$}\fi}

\newcommand{\EX}{\ifmmode{\xi}\else{$\xi$}\fi}
\newcommand{\EXF}{\ifmmode{\phi}\else{$\phi$}\fi}
\newcommand{\extend}[2]{#1\circ#2}

\newcommand{\acts}{\alpha}

\newcommand{\st}{\sigma}

\newcommand{\inter}[1]{
	\ifmmode{\left(\bigcap_{\mathcal{Q}\in#1}\mathcal{Q}\right)}
	\else{$\left(\bigcap_{\mathcal{Q}\in#1}\mathcal{Q}\right)$}
	\fi
}

\newcommand{\idSet}{\mathcal{I}}
\newcommand{\wSet}{\mathcal{W}}
\newcommand{\rdSet}{\mathcal{R}}
\newcommand{\srvSet}{\mathcal{S}}

\newcommand{\op}{\pi}

\newcommand{\rd}{\rho}
\newcommand{\wrt}{\omega}

\mathchardef\mhyphen="2D

\newcommand{\pr}{p}
\newcommand{\rdr}{r}

\newcommand{\srvr}{s}

\newcommand{\bef}{\rightarrow}

\newcommand{\vid}[1]{\ifmmode{\nu_{#1}}\else{$\nu_{#1}$}\fi}

\newcommand{\seen}{\ifmmode{seen}\else{$seen$}\fi}

\newcommand{\ABD}{{\sc ABD}}

\newcommand{\tg}[1]{tg_{#1}}

\newcommand{\maxts}[1]{\ifmmode{maxTS_{#1}}\else{$maxTS_{#1}$}\fi}
\newcommand{\maxtag}[1]{\ifmmode{maxTag_{#1}}\else{$maxTag_{#1}$}\fi}
\newcommand{\maxpair}[1]{\ifmmode{maxMPair_{#1}}\else{$maxMPair_{#1}$}\fi}
\newcommand{\mintag}[1]{\ifmmode{minTag_{#1}}\else{$minTag_{#1}$}\fi}
\newcommand{\maxps}{\ifmmode{maxPS}\else{$maxPS$}\fi}
\newcommand{\conftg}[1]{\ifmmode{confirmed_{#1}}\else{$confirmed_{#1}$}\fi}
\newcommand{\maxconftag}{\ifmmode{\ms{maxCT}}\else{$maxCT$}\fi}
\newenvironment{proof}{Proof}

\providecommand{\tsmp}[1]{ts_{#1}}

\graphicspath{ {sim_charts/} }

\newcommand{\ppp}[1]{\smallskip\noindent{\bf #1}}

\renewcommand{\ABD}{{\small ABD}}
\newcommand{\ABDmwmr}{{\small ABD-MW}}
\newcommand{\oram}{{\sc OhRam}}
\newcommand{\osam}{{\small\sc OhSam}}
\newcommand{\ohsam}{{\small\sc OhSam}}
\newcommand{\omam}{{\small\sc OhMam}}
\newcommand{\ohmam}{{\small\sc OhMam}}
\newcommand{\osamex}{{\small\sc OhSam'}}
\newcommand{\omamex}{{\small\sc OhMam'}}
\newcommand{\benchmark}{{\small\sc LB}}

\newcommand{\metrics}{{exchanges}}

\newcommand{\x}[1]{{\sc e}#1}
\newcommand{\ex}[1]{{\sc e}#1}

\newcommand{\sw}{{\small SWMR}}
\newcommand{\mw}{{\small MWMR}}

\begin{document}

\title{ Oh-RAM! One and a Half Round \\
Atomic Memory}

\author{Theophanis Hadjistasi
		 \thanks{University of Connecticut, Storrs CT, USA. Email: {\tt theo@uconn.edu, aas@engr.uconn.edu}}
		\and Nicolas Nicolaou
		\thanks{IMDEA Networks Institute, Madrid, Spain. Email: {\tt nicolas.nicolaou@imdea.org}}
		\and Alexander Schwarzmann$^*$}

\maketitle

\begin{abstract}

Implementing atomic read/write shared objects in a message-passing system
is an important problem in distributed computing.
Considering that communication is the most expensive 
resource, efficiency of read and write operations is assessed primarily in terms of the
needed communication and the associated latency. 
The seminal result of Attiya, Bar-Noy, and Dolev established that two communication round-trip
phases involving in total \emph{four} message exchanges are sufficient to implement atomic operations
when a majority of processors are correct. 
Subsequently it was shown by Dutta et al. that
one round involving \emph{two} communication exchanges is sufficient as long as the system adheres to certain
constraints with respect to crashes on the number of readers and writers in the system. 
It was also observed that three message exchanges are sufficient in some settings.


This work explores the possibility of devising algorithms 
where operations are able to complete in \emph{three} communication exchanges without imposing 
constraints on the number of participants, i.e.,
the aim is  \emph{One and half Round Atomic Memory}, hence the name \oram{}!
A recent result by Hadjistasi et al.
suggests that three-exchange implementations are \emph{impossible} in the 
MWMR (multi-writer/multi-reader) setting. 
This paper shows that this is achievable in the SWMR
(single-writer/multi-reader)
setting and also achievable for read operations
in the MWMR setting by ``sacrificing'' the performance of 
write operations. In particular,  
we present an atomic SWMR memory implementation,
where reads complete in \emph{three} 
and writes complete in \emph{two}  communication exchanges. 
Next, we provide an atomic MWMR memory implementation,
where reads involve \emph{three} and writes involve \emph{four} communication exchanges.
In light of the impossibility result 
these  algorithms are optimal in 
terms of the number of communication exchanges.
%
Both algorithms are then refined to allow 
some reads to complete in just \emph{two} communication exchanges. 
To evaluate these algorithms
we use the NS3 simulator and compare their performance 
in terms of operation latency. 
The algorithms are evaluated with different
topologies and operation loads.
\end{abstract}

%
%

\newpage

\section{Introduction}
\label{sec:oram:intro}
Emulating atomic \cite{Lamport79} (or linearizable \cite{HW90}) read/write objects in 
message-passing environments is an important problem in distributed computing. 
Atomicity is the 
most intuitive consistency semantic as it provides the illusion of a single-copy object 
that serializes all accesses such that each read operation returns the value of the latest
preceding write operation.
Solutions to this problem are complicated when the processors 
are failure-prone  and when the environment 
is asynchronous.
To cope with processor failures, distributed object 
implementations
use \emph{redundancy} by replicating the object 
at multiple network locations.
%
Replication introduces the problem of consistency 
because operations may access different object replicas
possibly containing obsolete values.

The seminal work of Attiya, Bar-Noy, and Dolev \cite{ABD96}
provided an algorithm, colloquially referred to as \ABD{}, 
that implements single-writer/multiple-reader ({\small SWMR}) atomic objects
in message-passing crash-prone asynchronous environments. 
The operations are ordered with the help
of logical \emph{timestamps} associated with each value. 
Here each operation is guaranteed to terminate as long as some majority of replica servers do not crash. 
Each write operation takes one communication round-trip phase, or round, involving 
\emph{two} communication exchanges and each read operation 
 takes two rounds involving in total
\emph{four} communication exchanges.
%
%
Subsequently, \cite{LS97} showed how to implement multi-writer/multiple-reader 
({\small MWMR})
atomic memory where both read and write operations involve two communication
round trips involving in total four communication exchanges.

The work by Dutta et al.~\cite{CDGL04} introduced a {\small SWMR} implementation
where both reads and writes involve a single round
consisting of \emph{two} communication exchanges.
Such an implementation
is called \emph{fast}, and it was shown that this is possible only when 
the number of readers $r$ is bounded with respect to the number of servers $s$
and the number of server failures $f$, viz. $r<\frac{s}{f}-2$.
%
An observation made in \cite{CDGL04} suggests that 
atomic memory may be implemented (using a max/min technique)
so that each read and write operation complete in \emph{three} communication exchanges. 
The authors did not elaborate on the inherent limitations that such a technique 
may impose on the distributed system. 

Subsequent works, e.g., \cite{EGMNS09,GNS09}, 
focused in relaxing 
the bound on the number of readers and writers in the service by proposing 
hybrid approaches where some operations complete 
in \emph{one} and others in \emph{two} rounds.
Tight bounds were provided in \cite{EGMNS09} on 
the number of rounds that {read} and {write}
operations require in the {\small MWMR} model.

A natural question arises whether one can devise 
implementations where all operations complete in at most \emph{three}
communication exchanges without imposing any restrictions 
on the numbers of participants in the service. 
A recent work by Hadjistasi, Nicolaou, and
Schwarzmann \cite{HNS2017} showed that such implementations
are impossible in the {\small MWMR} setting. It is not known
whether there is an {\small SWMR} implementation and whether
there exists some trade off that allows operations to complete
in three communication exchanges 
in the {\small MWMR} setting.

\begin{table}[t!h]
\begin{center}
{\small
\begin{tabular}{| c | c | c | c | c | c |}
\hline
  Model & Algorithm & Read Exchanges & Write Exchanges & Read Comm. & Write Comm.\\
  \hline\hline
  SWMR & ABD 
   & 4 & 2 & $4|\srvSet|$ & $2|\srvSet|$\\ 
  \hline 
  SWMR & \osam{}  & 3 & 2 & $|\srvSet|^{2} + 2|\srvSet|$ & $2|\srvSet|$\\
  \hline 
  SWMR & \osamex{}  & 2 or 3 & 2 & $|\srvSet|^{2} + 3|\srvSet|$ & $2|\srvSet|$\\
  \hline
  MWMR & ABD 
   & 4 & 4 & $4|\srvSet|$ & $4|\srvSet|$\\
  \hline
  MWMR & \omam{}  & 3 & 4 & $|\srvSet|^{2} + 2|\srvSet|$ & $4|\srvSet|$\\
  \hline
  MWMR & \omamex{}  & 2 or 3 & 4 & $|\srvSet|^{2} + 3|\srvSet|$ & $4|\srvSet|$\\
  \hline
\end{tabular}
}
\end{center}
\vspace{-\bigskipamount}
\label{table:oram:complexities}
\caption{Summary of communication exchanges and communication complexities.}\vspace{-10mm}
\end{table}

\ppp{Contributions.}
We focus on the gap between one-round and two-round algorithms
by presenting atomic memory algorithms where read 
operations can take ``one and a half rounds,'' i.e.,
complete in \emph{three} communication exchanges.
%
We also provide {\small SWMR} and {\small MWMR} algorithms
where read operations complete
in either \emph{two} or \emph{three} communication exchanges. 
We rigorously reason about the correctness of the algorithms.
To assess the practicality of these implementations we
simulate them and compare their performance.
Additional details are as follows.
%
\begin{enumerate}
	\item We present a new {\small SWMR} algorithm (\osam{}) for atomic objects in the asynchronous
	message-passing model with processor crashes.
	Write operations take \emph{two}
	communication exchanges and are similar to the write operations of \ABD{}.
	Read operations take \emph{three} communication exchanges: (1) the reader 
	sends a message to  servers, 
	(2) the servers share this information, 
	and (3)~once this is ``sufficiently'' done, servers reply to the reader. 
	A key idea of the algorithm is that the reader returns the value that is 
	associated with the \emph{minimum} timestamp (cf.~the observation in~\cite{CDGL04}). 
	The read operations are optimal in terms 
	of communication exchanges in light of \cite{HNS2017}.   
	(Section \ref{sec:osam:algorithm}.)
	
	\item We extend the {\small SWMR} algorithm to yield a {\small MWMR} algorithm (\omam{}). 
	In the new algorithm the write operations are more complicated, 
	taking \emph{four} communication exchanges
	(cf.~\cite{LS97}). Read operations complete as before in \emph{three} communication exchanges. 
	(Section \ref{sec:omam:algorithm}.)
	
	\item 
	We then present a revised  {\small SWMR} algorithm (\osamex{})
	and a revised {\small MWMR} algorithm (\omamex{}), where  read operations
	complete in either \emph{two} or \emph{three} communication exchanges.
	The original and the revised versions of each algorithm are presented for pedagogical reasons:
	for ease of understanding and reasoning about the algorithms.
	(Section \ref{sec:oram:extensions}.)
	
	\item We simulate our algorithms using the NS3 simulator and assess their performance
	under practical considerations. We note that the relative performance of our algorithms depends 
	on the simulation topologies and object server placement; this is another reason for presenting both versions of each algorithm.
	(Section \ref{sec:oram:simulation}.)
	
\end{enumerate}

The summary of  complexity results in comparison with ABD~\cite{ABD96}
is in Table~\ref{table:oram:complexities}. Improvements in the 
latency (in terms of the number of exchanges) are obtained in a trade-off with
communication complexity. 
{We note that increases in the communication complexity
need not necessarily have negative consequences in some practical settings, 
such as data centers, 
where servers communicate 
over high-bandwidth links.}

\section{Models and Definitions}
\label{sec:oram:model}
%
The system consists of a collection of crash-prone, {asynchronous processors} 
with unique identifiers from a totally-ordered set $\idSet$ 
partitioned into:
%
set $\wSet$ of writer identifiers,
set $\rdSet$ of reader identifiers, and
set $\srvSet$ of replica server identifiers
with each \textit{server}
maintaining a copy of the object.
Any subset of writers and readers, 
 and up to $f$ servers, $f < \frac{|\srvSet|}{2}$, may crash at any time.
%
Processors communicate by {exchanging messages} via {asynchronous} point-to-point
reliable channels; messages may be reordered. 
For convenience we use the term \emph{broadcast} as a shorthand
denoting sending point-to-point messages to multiple destinations.
\ppp{Executions.}
An algorithm $A$ is a collection of
processes, where 
process $A_\pr$ is assigned to processor $\pr\in\idSet$. The \emph{state} of 
processor $\pr$ is determined over a set of state variables, and the state of
$A$ is a vector that contains the state of each
process. 
Algorithm $A$ performs a \emph{step}, when some process $p$
(i) receives a message,
(ii) performs local computation,
(iii) sends a message.
Each such action  causes the state at $p$ to change.
An \emph{execution} is an 
alternating sequence of states and actions of $A$ starting with the initial state and 
ending in a state.
A process $\pr$ \emph{crashes} in an execution if
it stops taking steps; otherwise $\pr$ is \emph{correct}.

\ppp{Atomicity.} 
An implementation of a read or a write operation
contains an \emph{invocation} action 
(such as a call to a procedure) and a
\emph{response} action (such as a return from the procedure).
An operation $\op$ 
is \emph{complete} in an execution $\EX$, if $\EX$ contains 
both the invocation and the \emph{matching} response actions for $\op$; otherwise 
$\op$ is \emph{incomplete}. 
An execution 
is \emph{well formed} if any process 
invokes one operation at a time. 
We say that an operation $\op$ \emph{precedes} 
an operation $\op'$ in an execution $\EX$, denoted by $\op\bef\op'$, if the response step of $\op$ 
appears before the invocation step in $\op'$ in $\EX$. 
Two operations are
\emph{concurrent} if neither precedes the other. 
%
%
%
%
%
%
The correctness of an atomic read/write object implementation is defined in terms of  \emph{atomicity} (safety) 
and \emph{termination} (liveness) properties. Termination requires that any operation invoked by a correct
 process eventually completes.
Atomicity  is defined following \cite{Lynch1996}. 
For any execution $\EX$, 
if all invoked read and write operations are complete, then the
operations 
can be partially ordered by an ordering $\prec$, so that the following properties are satisfied:

\begin{description}
	\item[A1] The partial order $\prec$ is consistent with the external order of invocation and responses, that
	is, there do not exist operations $\op$ and $\op^{\prime}$, such that $\op$ completes before $\op^{\prime}$ starts, yet $\op^{\prime} \prec \op$.
	\item[A2] All write operations are totally ordered and every read operation is ordered with respect to all  writes.
	\item[A3] Every read operation returns the value of the last write preceding it in the partial order, and any read operation ordered before all writes returns the initial 	value of the object.
\end{description}

\newcommand{\phnum}{\mathfrak{p}}

\ppp{Efficiency and Communication Exchanges.} 
Efficiency of implementations is assessed in terms of
\emph{operation latency} and \emph{message complexity}.  
\emph{Latency} of each operation is 
determined by
the \emph{computation time} and the \emph{communication delays}.
{Computation time} accounts for the computation steps that 
the algorithm performs in each 
operation.  
{Communication delays} are measured in terms 
of \emph{communication $\metrics$}.
The protocol implementing each operation involves a collection of
sends (or broadcasts) of typed messages and the corresponding receives.
\emph{Communication exchange} within an execution of an operation
is the set of sends and receives for the specific message type
within the protocol. 
Note that using this definition, traditional implementations in the style of \ABD{}
are structured in terms of \emph{rounds}, cf. \cite{ABD96,GNS09},
 where each round consists
of two communication exchanges, the first, a broadcast, is initiated by the process executing
an operation, and the second, a convergecast, consists of responses to the initiator.
We refer to the $i^{th}$ exchange using the notation \x{i}. 
%
%
The number of messages that a process expects during a convergecast 
depends on the implementation. 
%
\emph{Message complexity} measures 
the {worst-case} total number of messages exchanged during an operation.


\section{SWMR Algorithm \osam{}}
\label{sec:osam:algorithm}

We now present our \sw{} algorithm \ohsam{}:
\emph{{\sf O}ne and a {\sf H}alf round {\sf S}ingle-writer {\sf A}tomic {\sf M}emory}. 
The write operations
are fast, that is, they take \emph{two}
	communication exchanges to complete (similarly to \ABD{}~\cite{ABD96}).
We show that atomic operations do not need to involve complete communication round
trips between clients and servers.
In particular, we allow \emph{server-to-server} communication and we devise 
read operations that take \emph{three} communication exchanges using the following communication pattern: 
	exchange \ex{1} the reader sends message to the participating servers, 
	in exchange \ex{2} each server that receives the request it then \emph{relays} the request to all the servers, 
	and 
	once a server receives the relay for a particular read from a majority of servers, it 
	replies to the requesting reader forming exchange \ex{3}. 
	The read completes once the invoker collects a majority of acknowledgment replies.
	A key idea of the algorithm is that the reader returns the value that is 
	associated with the \emph{minimum} timestamp. 
	In particular, while the replica servers update their local value to 
	the associated with the \emph{maximum} timestamp received, the reader 
	returns the value associated with the \emph{minimum} timestamp discovered
	in the set of the received acknowledgment messages.
	The code is given in Algorithm~\ref{alg:osam}.
	We now give the details of the protocols; in
	referring to the numbered lines of code we use the prefix ``L'' to stand for ``line''.

Counter variables $read\_op$, $operations$ and $relays$ are used to help processes identify 
``new'' read and write operations, and distinguish ``fresh'' from ``stale'' messages 
(since messages can be reordered). 
The value of the object and its associated timestamp, as known by each process, are stored in variables $v$ and $ts$ respectively. 
Set $rAck$,
 at each reader $\rdr_i$, stores all the received acknowledgment messages. 
 Variable $minTS$ holds the minimum timestamp discovered in the set of the received acknowledgment messages $rAck$.
Below we provide a brief description of the protocol of each participant of the service.

\noindent{\bf Writer Protocol.} 
Writer $w$ increments its local timestamp  $ts$  
and broadcasts a \act{writeRequest} message 
to all the participating servers $\srvSet$ (L\ref{line:osam:client:inctimestamp}-\ref{line:osam:client:wbdcast}). 
Once the writer receives replies from at least a majority of servers, ${|\srvSet|}/{2} +1$, the operation completes (L\ref{line:osam:client:waitwritemajority}-\ref{line:osam:client:waitfresh}).

\noindent{\bf Reader Protocol.} 
When a read process $\rdr$ invokes a read operation
it first monotonically increases its local read operation counter $read\_op$  and empties the 
set of the received acknowledgment messages, $rAck$ (L\ref{line:osam:client:increadop}). 
Then, it creates a 
$\tup{{ \sf readRequest}, r, read\_op}$
\act{readRequest} message in which it encloses its id and local read counter 
and it broadcasts this request message to all the participating servers $\srvSet$, forming exchange \ex{1} (L\ref{line:osam:client:bdcast}).
It then waits to collect at least ${|\srvSet|}/{2} +1$ messages from servers. 
While collecting \act{readAck} messages from exchange \ex{3}, 
reader $\rdr$ discards any delayed messages from previous operations due to asynchrony.
When ``fresh'' messages are collected from a majority of servers, 
then the reader returns the value $v$ associated with the \emph{minimum} timestamp, $minTS$, 
among the set of the received acknowledgment messages, $rAck$
(L\ref{line:osam:client:reachmajority:start}-\ref{line:osam:client:reachmajority:end}). 


\begin{algorithm}[!ht]
\caption{\small Reader, Writer, and Server Protocols for SWMR algorithm \osam{}}
\label{alg:osam}
\begin{multicols}{2}
{\sf\footnotesize
\begin{algorithmic}[1]
\State {At each reader $r$}
\State \textbf{Variables:}
\State $ts \in \mathbb{N}^{+}$, $minTS \in \mathbb{N}^{+}$, $v \in V$
\State $read\_op \in \mathbb{N}^{+}$, $rAck \subseteq \srvSet \times M$
\State \textbf{Initialization:}
\State $ts \leftarrow 0$, $minTS \leftarrow 0$, $v \leftarrow \perp$, $read\_op \leftarrow 0$ 
\Function{Read}{}
	\State $read\_op \leftarrow read\_op + 1$ 					           \label{line:osam:client:increadop}
	\State $rAck \leftarrow \emptyset$
	\State \textbf{broadcast} $(\tup{{\sf readRequest}, r, read\_op})$ to $\srvSet$ \label{line:osam:client:bdcast}
	\State \textbf{wait until} $(|rAck| = {|\srvSet|}/{2} +1)$	
	\State $minTS \leftarrow min\{m.ts^{\prime}| m \in rAck \}$ \label{line:osam:client:reachmajority:start}
	\State $v \gets \{m.val$ $|$ $m\in rAck \land m.ts'=minTS \}$
	\State \textbf{return$(v)$}			 \label{line:osam:client:reachmajority:end}
\EndFunction
\newline
\State \textbf{Upon receive} $m$ from $s$
 	 \If{$m.read\_op = read\_op$}
 	 \State $rAck \gets rAck \cup \{(s,m)\}$
\EndIf
\newline
\State {At writer $w$}
\State \textbf{Variables:}
\State $ts \in \mathbb{N}^{+}$, $v \in V$, $wAck \subseteq \srvSet \times M$
\State \textbf{Initialization:}
\State $ts \leftarrow 0$, $v \leftarrow \perp$
\Function{Write}{$val: input$}
	\State $(ts,v) \leftarrow (ts+1, val)$        \label{line:osam:client:inctimestamp}
	\State $wAck \leftarrow \emptyset$
	\State \textbf{broadcast} $(\tup{{\sf writeRequest}, ts, v, w})$ to $\srvSet$ \label{line:osam:client:wbdcast}
	\State \textbf{wait until} $(|wAck| = {|\srvSet|}/{2} +1)$	  \label{line:osam:client:waitwritemajority}  \label{line:osam:client:waitfresh}
	\State \textbf{return}
\EndFunction
\newline
\State \textbf{Upon receive} $m$ from $s$
 	 \If{$m.ts = ts$}
 	 \State $wAck \gets wAck \cup \{(s,m)\}$
\EndIf
\newline
\pagebreak
\State {At server $s_i$}
\State \textbf{Variables:}
\State $ts \in \mathbb{N}^{+}$, $v \in V$
\State $operations[1..|\rdSet|$$+$$1]$, $relays[1..|\rdSet|$$+$$1]$ : array of int
\State \textbf{Initialization:}
\State $ts \leftarrow 0$, $v \leftarrow \perp$
\State $operations[i] \leftarrow 0$ for $i \in \rdSet$, $relays[i] \leftarrow 0$ for $i \in \rdSet$
\newline
\State \textbf{Upon receive}{$(\tup{{\sf readRequest}, r, read\_op})$}{} 	\label{line:osam:srv:readrcv:start}
	\State ~ ~ \textbf{broadcast}$(\tup{{\sf readRelay}, ts, v, r, read\_op, s_i})$ to  $\srvSet$\label{line:osam:srv:readrcv:end} \label{line:osam:srv:readrequestbdcast}
\newline
\State \textbf{Upon receive}$(\tup{{\sf writeRequest}, ts^{\prime}, v^{\prime}, w})$
	\If{$(ts < ts^{\prime})$}			\label{line:osam:srv:writeupdate:start}
		\State $(ts,v)$ $\leftarrow$ $(ts^{\prime}, v^{\prime})$
	\EndIf														\label{line:osam:srv:writeupdate:end}
	\State \textbf{send} $(\tup{ {\sf writeAck},ts, v,s_i})$ to $w$ \label{line:osam:srv:ackclient}
\newline
\State \textbf{Upon receive}$(\tup{ {\sf readRelay}, ts^{\prime}, v^{\prime}, r, read\_op, s_i})$
	~ \If{ $(ts < ts^{\prime})$}						\label{line:osam:srv:relayupdate:start}
		\State $(ts,v)$ $\leftarrow$ $(ts^{\prime},v^{\prime})$
	\EndIf									\label{line:osam:srv:relayupdate:end}
	~ \If{ $(operations[r] < read\_op)$}  				 \label{line:osam:srv:checkreadop:start} \label{line:osam:srv:checkrdop}
		\State $ operations[r] \leftarrow read\_op $
		\State $ relays[r] \leftarrow 0 $.
	\EndIf 									\label{line:osam:srv:checkreadop:end}
	~ \If{$(operations[r] = read\_op)$} 				\label{line:osam:srv:increlays:start}
		\State $relays[r] \leftarrow relays[r] + 1$ 
	\EndIf									\label{line:osam:srv:increlays:end}	
	~ \If{$(relays[r] = {|\srvSet|}/{2} +1)$}			\label{line:osam:srv:reachmajority:start}
		\State \textbf{send} $(\tup{ {\sf readAck},ts, v, read\_op, s_i})$ to $r$
	\EndIf \label{line:osam:srv:reachmajority:end}
\end{algorithmic}
}
\end{multicols}\vspace*{-\bigskipamount}
\end{algorithm}

\noindent{\bf Server Protocol.} 
Each server $\srvr \in \srvSet$ expects three types of messages:

$(1)$ Upon receiving a $\tup{{\sf readRequest}, r, read\_op}$ message
the server creates a \act{readRelay} message, containing its $ts$, $v$, and its id $s$,
and it broadcasts it to all the servers $\srvSet$
(L\ref{line:osam:srv:readrcv:start}-\ref{line:osam:srv:readrcv:end}).

$(2)$ Upon receiving message $\tup{{\sf readRelay}, ts^{\prime}, v^{\prime}, r, read\_op}$ 
server $\srvr$ compares its local timestamp $ts$ with $ts^{\prime}$ enclosed in the message. 
If $ts < ts^{\prime}$, 
then $\srvr$ sets 
its local timestamp and value
to those enclosed 
in the message
(L\ref{line:osam:srv:relayupdate:start}-\ref{line:osam:srv:relayupdate:end}). 
In any other case, no updates are taking place. 
As a next step
$\srvr$ checks if the received \act{readRelay}
message marks a new read operation by $r$.
This is achieved by checking if reader's $r$ operation counter is newer than the local one, 
 i.e., $read\_op > operations[r]$ (L\ref{line:osam:srv:checkrdop}).
If this holds, then $\srvr$: 
a) sets its local read operation  counter for reader $r$ to be equal to 
the received counter, i.e., $operations[r] = read\_op$; 
and 
b) re-initializes the relay counter for $r$  
to zero, i.e., $relays[r] = 0$ (L\ref{line:osam:srv:checkreadop:start}-\ref{line:osam:srv:checkreadop:end}).
Server $\srvr$ also updates the number of collected \act{readRelay} messages regarding the read request created by reader $r$
(L\ref{line:osam:srv:increlays:start}-\ref{line:osam:srv:increlays:end}). 
When $\srvr$ receives 
$\tup{{\sf readRelay }, ts, v,  read\_op, s_i}$  messages
from a majority of servers, 
it creates a $\tup{{ \sf readAck},ts, v, read\_op, s}$ message 
in which it encloses
its local timestamp and value,
its id, and 
the reader's operation counter 
and sends it to the requesting reader $r$ 
(L\ref{line:osam:srv:reachmajority:start}-\ref{line:osam:srv:reachmajority:end}).

$(3)$ Upon receiving message $\tup{{\sf writeRequest}, ts^{\prime}, v^{\prime}, w}$ 
server $\srvr$ compares its local timestamp $ts$ with the received one, $ts^{\prime}$. 
If $ts < ts^{\prime}$, then the server sets its local timestamp and value to be equal to those 
in the received message
(L\ref{line:osam:srv:writeupdate:start}-\ref{line:osam:srv:writeupdate:end}). 
In any other case, no updates are taking place. 
Finally, the server always sends an acknowledgement, \act{writeAck},
message to the requesting writer (L\ref{line:osam:srv:ackclient}).  

\subsection{Correctness.}
\label{sec:osam:correct}

To prove correctness of algorithm \osam{} 
we reason about its \emph{liveness} (termination) and \emph{atomicity} (safety).

\noindent{\bf Liveness.} 
Termination holds
with respect to our failure model:
up to $f$ servers may fail, where $f < |\srvSet|/2$
and each operation waits for messages from some majority of servers.
We now give more detail on how each operation satisfies \emph{liveness}.

\noindent{\it Write Operation.} 
Per algorithm \ohsam{}, writer $w$ creates a \act{writeRequest} message and 
then it broadcasts it to all  servers in exchange \ex{1} (L\ref{line:osam:client:wbdcast}). 
Writer $w$ then waits for \act{writeAck} messages from a majority of servers from \ex{2}
(L\ref{line:osam:client:waitwritemajority}-\ref{line:osam:client:waitfresh}). 
Since in our failure model up to $f < \frac{|\srvSet|}{2}$ servers may crash,  
writer $w$ collects \act{writeAck} messages form a majority of live servers during \ex{2} and 
the write operation $\omega$ terminates.
\smallskip

\noindent{\it Read Operation.} 
The reader $\rdr$ begins by broadcasting a \act{readRequest} message all the servers  forming exchange \ex{1}. 
Since $f < \frac{|\srvSet|}{2}$, then at least a majority of servers receives the \act{readRequest} message sent in \ex{1}. 
Any server $\srvr$ that receives this message it then broadcasts a \act{readRelay} message to all the servers, forming \ex{2},
(L\ref{line:osam:srv:readrcv:start}-\ref{line:osam:srv:readrcv:end}),
and no server ever discards any incoming \act{readRelay} messages. 
Any server, whether it is aware or not of the \act{readRequest}, 
always keeps a record of the incoming \act{readRelay} messages 
and takes action as if it is aware of the \act{readRequest}. 
The only difference between server $\srvr_i$ that received a \act{readRequest} message 
and server $\srvr_k$ that did not, 
is that $\srvr_i$ is able to broadcast a \act{readRelay} message, 
and $\srvr_k$  broadcasts a \act{readRelay} message 
when it receives the corresponding \act{readRequest} message 
(L\ref{line:osam:srv:readrcv:start}-\ref{line:osam:srv:readrcv:end}).   
Each non-failed server receives \act{readRelay} messages from a majority of servers during \ex{2}
and sends a \act{readAck} message to the requesting reader 
$\rdr$ at \ex{3} (L\ref{line:osam:srv:increlays:start}-\ref{line:osam:srv:increlays:end}). 
Therefore, reader $\rdr$ collects \act{readAck} messages from a majority of servers during \ex{3}, 
and the read operation terminates
(L\ref{line:osam:client:reachmajority:start}-\ref{line:osam:client:reachmajority:end}).

Based on the above, 
it is always  the case that acknowledgment messages 
{{\sf{readAck}}} and {{\sf{writeAck}}} are collected from at least a majority of servers 
in any read and write operation, thus ensuring \emph{liveness}. 
\smallskip

\paragraph{Atomicity.} To prove atomicity we order the operations 
with respect to timestamps written and returned.
More precisely, for each execution $\EX$ of the algorithm there must
exist a partial order $\prec$ on the operations in on the set of completed operations  $\Pi$ that satisfy conditions A1, A2, and A3
as given in Section~\ref{sec:oram:model}.
Let $ts_\pi$  
be the value of the timestamp at the completion of $\pi$ 
when  $\pi$ is a write, 
and the timestamp computed as the \emph{maximum} $ts$ at the completion of a read operation $\pi$. 
With this, we denote the partial order on operations as follows. 
For two operations $\pi_1$ and $\pi_2$, when $\pi_1$ is any operation and $\pi_2$ is a write, 
we let $\pi_1 \prec \pi_2$ if $ts_{\pi_1} < ts_{\pi_2}$.
For two operations $\pi_1$ and $\pi_2$,
when $\pi_1$ is a write and $\pi_2$ is a read we let $\pi_1 \prec \pi_2$ if $ts_{\pi_1} \leq ts_{\pi_2}$.
The rest of the order is established by transitivity and reads with the same timestamps are not ordered.
We now state and prove the following lemmas.

It is easy to see that the $ts$ variable in each server $s$ is monotonically increasing. This leads
to the following lemma.

\begin{lemma} 
\label{lem:osam:srv:monotonic}
In any execution $\EX$ of \osam{}, the variable $ts$ maintained by any server $\srvr$ in the system is non-negative and monotonically increasing.
\end{lemma}

\begin{Proof}
When a server $\srvr$ receives a timestamp $ts$ then $\srvr$ updates its 
local timestamp $ts_s$ if and only if $ts > ts_s$ (L\ref{line:osam:srv:writeupdate:start}-\ref{line:osam:srv:writeupdate:end} 
and L\ref{line:osam:srv:relayupdate:start}-\ref{line:osam:srv:relayupdate:end}). 
Thus the local timestamp of the server monotonically increases
and the lemma follows. 
\end{Proof}

Next, we show that if a read operation $\rd_2$ succeeds read operation $\rd_1$, 
then $\rd_2$ always returns a value at least as recent as the one returned by $\rd_1$.

\begin{lemma}
\label{lem:osam:read-after-read}
In any execution $\EX$ of \osam{}, if $\rd_1$ and $\rd_2$ 
are two read operations such that $\rd_1$ precedes $\rd_2$, i.e., 
$\rd_1 \to \rd_2$, and $\rd_1$ returns the value for timestamp $ts_1$, then $\rd_2$ returns 
the value for timestamp $ts_2 \geq ts_1$.
\end{lemma}

\begin{Proof} 
Let the two operations $\rd_1$ and $\rd_2$ be invoked by processes with identifiers
$\rdr_1$ and $\rdr_2$ respectively (not necessarily different). Also, let $RSet_1$ and 
$RSet_2$ be the sets of servers that sent a {{\sf{readAck}}} message to $\rdr_1$ and 
$\rdr_2$ during $\rd_1$ and $\rd_2$.

Assume by contradiction
that read operations $\rd_1$ and $\rd_2$ exist such that $\rd_2$ succeeds $\rd_1$, 
i.e., $\rd_1 \to \rd_2$, and the operation $\rd_2$ returns a timestamp $ts_2$ 
that is smaller than the $ts_1$ returned by $\rd_1$, i.e., $ts_2 < ts_1$. 
According to our algorithm,  
$\rd_2$ returns a timestamp $ts_2$ that is smaller than 
the minimum timestamp received by $\rd_1$, i.e., $ts_1$, if 
$\rd_2$ obtains  $ts_2$ and $v$ in the {{\sf{readAck}}} message of some server 
$\srvr_x\in RSet_2$, and $ts_2$ is the minimum timestamp received by $\rd_2$.

Let us examine if $\srvr_x$  
replies with $ts^{\prime}$ and $v^{\prime}$ to $\rd_1$, 
i.e., $\srvr_x\in RSet_1$. By 
Lemma \ref{lem:osam:srv:monotonic}, and since $\rd_1\bef\rd_2$, then 
it must be the case that $ts^{\prime}\leq ts_2$.
According to our assumption $ts_1>ts_2$, and since $ts_1$ is the 
smallest timestamp sent to $\rd_1$ by any server in $RSet_1$, then it follows that $\rdr_1$ does 
not receive the {{\sf{readAck}}} message from $\srvr_x$, and hence $\srvr_x\notin RSet_1$. 

Now let us examine the actions of the server $\srvr_x$. From the algorithm, 
server $\srvr_x$ collects {\sf{readRelay}} messages from a majority 
of servers in $\srvSet$ before sending a {{\sf{readAck}}} message to $\rd_2$
(L\ref{line:osam:srv:reachmajority:start}-\ref{line:osam:srv:reachmajority:end}).  
Let $RRSet_{\srvr_x}$ denote the set of servers that sent {\sf{readRelay}} to $\srvr_x$.
Since, both $RRSet_{\srvr_x}$ and $RSet_1$ contain some majority of the servers 
then it follows that $RRSet_{\srvr_x}\cap RSet_1\neq\emptyset$.
 
Thus there exists a server $\srvr_i\in RRSet_{\srvr_x}\cap RSet_1$, which sent 
(i) a {\sf{readAck}} to $\rdr_1$ for $\rd_1$, and (ii) a {\sf{readRelay}} to 
$\srvr_x$ during $\rd_2$. Note that $\srvr_i$ sends a {\sf{readRelay}} for $\rd_2$ only after
it receives a read request from $\rd_2$ (L\ref{line:osam:srv:readrcv:start}-\ref{line:osam:srv:readrcv:end}).
Since $\rd_1\bef\rd_2$, then it follows that $\srvr_i$ sent the 
{\sf{readAck}} to $\rd_1$ before sending the {\sf{readRelay}} to $\srvr_x$. By 
Lemma \ref{lem:osam:srv:monotonic}, if $\srvr_i$ attaches a timestamp $ts_{s_i}$ 
in the {\sf{readAck}} to $\rd_1$, then $\srvr_i$ attaches a timestamp $\tsmp{\srvr_i}'$ 
in the {\sf{readRelay}} message to $\srvr_x$, such that  $\tsmp{\srvr_i}'\geq\tsmp{\srvr_i}$. 
Since $\tsmp{1}$ is the minimum timestamp received by $\rd_1$, then $\tsmp{\srvr_i}\geq \tsmp{1}$,
and hence $\tsmp{\srvr_i}'\geq \tsmp{1}$ as well. By Lemma \ref{lem:osam:srv:monotonic}, and since 
$\srvr_x$ receives the {\sf{readRelay}} message from $\srvr_i$ before sending a {\sf{readAck}}
to $\rd_2$, it follows that $\srvr_x$ sends a timestamp $\tsmp{2}\geq\tsmp{\srvr_i}'$.
Thus, $\tsmp{2}\geq\tsmp{1}$ and this contradicts our initial assumption. 
\end{Proof}

{The next lemma shows that any read operation following a write operation  receives {\sf{readAck}} messages from servers where each included timestamp is at least as large as the one returned by the complete write operation.}

\begin{lemma}
\label{lem:osam:read-after-write-step1}
In any execution $\EX$ of \osam{}, if a read operation $\rd$ succeeds a write 
operation $\omega$ that writes  $ts$ and $v$, i.e., $\omega \bef \rd$, and 
receives {\sf{readAck}} messages from a majority of servers $RSet$, then 
each $\srvr\in RSet$ sends a {\sf{readAck}} message to $\rd$ with a timestamp $\tsmp{s}$ s.t. $\tsmp{s} \geq ts$. 
\end{lemma}

\begin{Proof}
Let $WSet$ be the set of servers that send a {\sf{writeAck}} message in $\omega$ and let $RRSet$ be the set of servers that send {\sf{readRelay}} messages to server $\srvr$.  

By Lemma \ref{lem:osam:srv:monotonic}, if a server $\srvr$ receives 
timestamp $ts$ from 
process $\pr$,
then $\srvr$ includes timestamp $ts^{\prime}$ s.t. $ts^{\prime} \geq ts$ in any subsequent message. 
This, means that 
every server in $WSet$, sends a {\sf{writeAck}} message to $\omega$ with a timestamp greater or equal to $ts$. Hence, every server $\srvr_x \in WSet$ has timestamp $\tsmp{\srvr_x} \geq ts$. Let us now examine timestamp $\tsmp{s}$ that server $\srvr\in RSet$ sends in read operation $\rd$.  

Before server $\srvr$ sends a {\sf{readAck}} message in $\rd$, 
it must receive {\sf{readRelay}} messages from the majority of servers, $RRSet$ (L\ref{line:osam:srv:reachmajority:start}-\ref{line:osam:srv:reachmajority:end}). 
Since both $WSet$ and $RRSet$ contain a majority of servers, 
then $WSet\cap RRSet\neq\emptyset$. 
By Lemma \ref{lem:osam:srv:monotonic}, any server $\srvr_x \in WSet\cap RRSet$ has a timestamp $\tsmp{s_x}$ s.t. $\tsmp{s_x} \geq ts$.
Since server $\srvr_x \in RRSet$ and from the algorithm, server's $\srvr$ timestamp is always updated
 to the highest timestamp it receives (L\ref{line:osam:srv:relayupdate:start}-\ref{line:osam:srv:relayupdate:end}), 
 then when server $\srvr$ receives the message from $\srvr_x$, it updates its timestamp $\tsmp{s}$ s.t. $\tsmp{s} \geq \tsmp{s_x}$.
%
Thus, by Lemma \ref{lem:osam:srv:monotonic}, each $\srvr\in RSet$ sends a {\sf{readAck}} (L\ref{line:osam:srv:reachmajority:start}-\ref{line:osam:srv:reachmajority:end})
 in $\rd$ with a timestamp $\tsmp{s}$ s.t. $\tsmp{s} \geq \tsmp{s_x} \geq ts$. Therefore,  $\tsmp{s} \geq ts$ holds and the lemma follows.
\end{Proof}

{Next show that if a read operation succeeds a write operation, then it returns a value at least as recent as the one that was written.}

\begin{lemma}
\label{lem:osam:read-after-write}
In any execution $\EX$ of \osam{}, if a read $\rd$ succeeds a write operation $\omega$ that writes timestamp $ts$, i.e. $\omega \to \rd$, and returns a timestamp $ts'$, then $ts^{\prime} \geq ts$.      
\end{lemma}

\begin{Proof}
Suppose that read operation $\rd$ receives {\sf{readAck}} messages from a majority of servers $RSet$. 
By lines \ref{line:osam:client:reachmajority:start}-\ref{line:osam:client:reachmajority:end} of the algorithm, it follows that $\rd$ 
decides on the minimum timestamp, $ts^{\prime}=ts\_min$, among all the timestamps in the {\sf{readAck}} messages 
of the servers in $RSet$.
From Lemma \ref{lem:osam:read-after-write-step1},  $ts\_min \geq ts$ holds, where $ts$ is the timestamp written by the last complete write operation $\omega$, then $ts^{\prime} = ts\_min \geq ts$ also holds. Therefore, $ts^{\prime} \geq ts$ holds and the lemma follows.
\end{Proof}

\begin{theorem}
Algorithm {\small \sc \osam{}} implements an atomic {\small SWMR} object.
\end{theorem}

\begin{Proof} 
We now use the lemmas stated above and the operations order definition
to reason about each of the three \emph{atomicity} conditions A1, A2 and A3. 

\ppp{A1}  For any $\op_1,\op_2\in\Pi$ such that $\op_1\bef\op_2$, it cannot be that $\op_2\prec\op_1$.
		
\noindent 	When the two operations $\op_1$ and $\op_2$ are reads and $\op_1\bef\op_2$ holds, then	
				from Lemma~\ref{lem:osam:read-after-read}
				it follows that the timestamp returned from $\op_2$ is always greater or equal to the one returned from $\op_1$,
				$ts_{\op_2} \geq ts_{\op_1}$.
				If $ts_{\op_2} > ts_{\op_1}$ then by the ordering definition $\op_1\prec\op_2$ is satisfied.
				When $ts_{\op_2} = ts_{\op_1}$ then the ordering is not defined, thus it cannot be the case that 
				$\op_2\prec\op_1$.
				If $\op_2$ is a write, the sole writer generates a new timestamp by
				incrementing the largest timestamp in the system. 
				By well-formedness (see Section \ref{sec:oram:model}), 
					any timestamp generated by the writer for any write operation that 
				precedes $\op_2$ must be smaller than $ts_{\op_2}$.
				Since $\op_1\bef\op_2$, then it holds that $ts_{\op_1} < ts_{\op_2} $. Hence, by the ordering
				definition it cannot be the case that $\op_2\prec\op_1$.
				Lastly, when $\op_2$ is a read and $\op_1$ a write and $\op_1\bef\op_2$ holds, 
				then from Lemmas~\ref{lem:osam:read-after-write-step1} and \ref{lem:osam:read-after-write}
				it follows that $ts_{\op_2} \geq ts_{\op_1}$.
				By the ordering definition, it cannot hold that $\op_2\prec\op_1$ in this case either.

\ppp{A2} For any write $\wrt\in\Pi$ and any operation $\op\in\Pi$, then either $\wrt\prec\op$ or $\op\prec\wrt$.

\noindent	If the timestamp returned from $\wrt$ is greater than the one returned from $\op$,
				i.e. $ts_{\wrt} > ts_{\op}$,
				 then $\op\prec\wrt$ follows directly.
				Similartly, if $ts_{\wrt} < ts_{\pi}$ holds, then $\wrt\prec\op$ follows.
				If $ts_{\wrt} = ts_{\pi}$, then it must be that $\op$ is a read 
				and 
				$\op$ discovered $ts_{\wrt}$ as the \emph{minimum} timestamp in at least a majority of servers.	
				Thus, $\wrt\prec\op$ follows.
			
\ppp{A3} Every read operation returns the value of the last write preceding it according to $\prec$ (or the initial value if there is no such write).
	
\noindent	Let $\wrt$ be the last write preceding read $\rho$.
				From our definition it follows that $ts_{\rho} \geq ts_{\wrt}$.
		    		If $ts_{\rho} = ts_{\wrt}$, then $\rho$ returned
		   		 the value written by $\wrt$ on a majority of servers.
		    		If $ts_{\rho} > ts_{\wrt}$, then it means that $\rho$ obtained a larger 
		    		timestamp. However, the larger timestamp can only be originating from a write
		    		that succeeds $\wrt$, thus $\wrt$ is not the preceding write and this cannot be the case.
		    		Lastly, if $ts_{\rho} = 0$, no preceding writes exist, and $\rho$ returns the initial value. 

\end{Proof}

Having shown liveness and atomicity of algorithm \osam{} the result follows.

\subsection{Performance.}
\label{sec:osam:performance}

We now assess the performance of \osam{} in terms of 
(i) latency of read and write operations as measured by the number of communication exchanges, and 
(ii) the message complexity of read and write operations.

In brief, for algorithm \osam{} write operations take {2} \metrics{} and
read operations take {3} communication \metrics{}.
The (worst case) message complexity of read operations
is $|\srvSet|^{2} + 2|\srvSet|$
and the (worst case)  message complexity of write operations
is $2|\srvSet|$.
This follows directly from the structure of the algorithm.
We now give additional details.

\ppp{Operation Latency.} ~
\emph{Write operation latency:} 
According to algorithm \osam{}, writer $w$ sends a {\sf{writeRequest}} message 
to all the servers in exchange \x{1},
and, awaits for {\sf{writeAck}} messages from at least a majority of servers in exchange \x{2}. 
Once the {\sf{writeAck}} messages are received, no further communication is required and the write operation
terminates. Therefore, any write operation  consists of \emph{2} communication exchanges. 

\emph{Read operation latency:} 
A reader sends a {\sf{readRequest}} message to all the servers
in the first communication exchange \x{1}. 
Once a server receives a {\sf{readRequest}} message, 
it broadcasts a {\sf{readRelay}} message to all the servers
in exchange \x{2}. 
Any active servers now await {\sf{readRelay}} messages from at least a majority of servers,
and then, the servers send a {\sf{readAck}} message to the reader
during communication exchange \x{3}. 
We note that servers do not reply  to any incoming {\sf{readRelay}} messages. 
Thus, a read operation consists of \emph{3} communication exchanges.

\ppp{Message Complexity.}
We measure operation message complexity as the worst case number of exchanged messages 
in each read and write operation. 
The worst case number of messages corresponds to failure-free executions where all
participants send messages to all destinations according to the protocols. 

\emph{Write operation:} 
A single write operation in algorithm \osam{} takes \emph{2} communication exchanges.
In the first exchange \x{1}, the writer sends a {\sf{writeRequest}} message to all the servers in $\srvSet$.
The second exchange \x{2}, occurs when all servers in $\srvSet$ send a {\sf{writeAck}} message 
to the writer. 
Thus, at most $2|\srvSet|$ messages are exchanged in a write operation.

\emph{Read operation:} 
Read operations take \emph{3} communication exchanges. 
Exchange \x{1} occurs when a reader sends a {\sf{readRequest}} message to all the servers in $\srvSet$. 
Exchange \x{2} occurs when servers in $\srvSet$ send a {\sf{readRelay}} message
to all the servers in $\srvSet$. 
The last exchange, \x{3}, occurs when servers in $\srvSet$ send a {\sf{readAck}} message 
to the requesting reader. 
Therefore, $|\srvSet|^{2} + 2|\srvSet|$ messages are exchanged during a read operation.

\section{MWMR Algorithm \omam{}}
\label{sec:omam:algorithm}

We seek a solution for the \mw{} setting that 
involves \emph{three} communications exchanges per read operation and
\emph{four} exchanges per write operation. 
We now present our \mw{} algorithm \omam{}:
\emph{{\sf O}ne and a {\sf h}alf round {\sf M}ulti-writer {\sf A}tomic {\sf M}emory}. 
To impose an ordering on the values 
written by the writers we 
associate each value with a tag \emph{tg} defined as the pair $\tup{ts,id}$, where 
$ts$ is a timestamp and \emph{id} is the identifier of a writer.
Tags are ordered lexicographically (cf.~\cite{LS97}). 
%
The read protocol is identical to the one presented in section \ref{sec:osam} for algorithm \osam{} 
(except that tags are used instead of timestamps). 
Thus, for algorithm \omam{} 
we briefly describe only
the protocols that differ, and that is, the writer and server protocols.


\begin{algorithm}[H]
\caption{\small Reader, Writer and Server Protocols for MWMR algorithm \omam{}}
\label{alg:omam}
\begin{multicols}{2}
{\sf\footnotesize
\begin{algorithmic}[1]
\makeatletter\setcounter{ALG@line}{54}\makeatother
\State \textit{At each reader $r$}
\State \textbf{Components:}
\State $tg \in \tup{\mathbb{N}^{+},\idSet}$, $minTAG \in \tup{\mathbb{N}^{+},\mathbb{N}^{+}}$
\State $v \in V$, $read\_op \in \mathbb{N}^{+}$, $rAck \subseteq \srvSet \times M$
\State \textbf{Initialization:}
\State $tg \leftarrow \tup{0,r}$,  $minTAG \leftarrow \tup{0,0}$
\State $v \leftarrow \perp$, $read\_op \leftarrow 0$
\Function{Read}{}
	\State $read\_op \leftarrow read\_op + 1$. 					      \label{line:omam:client:inc_read_op}
	\State $rAck \leftarrow \emptyset$
	\State \textbf{broadcast} $({\sf readRequest}, r, read\_op)$ to $\srvSet$  \label{line:omam:client:read_bdcast}
	\State \textbf{wait until} $(|rAck| = {|\srvSet|}/{2} +1)$	 \label{line:omam:client:wait_majority:start} \label{line:omam:client:wait_majority:end}
	\State $minTAG \leftarrow min(\{m.tg^{\prime} | m \in rAck \})$ \label{line:omam:client:reachmajority:start}
	\State $v \gets  \{ m.val$ $|$ $m \in rAck \land m.tg^{\prime}=minTAG\}$
	\State \textbf{return(v)}			\label{line:omam:client:reachmajority:end}
\EndFunction
\newline
\State \textbf{Upon receive} $m$ from $s$
 	 \If{$m.read\_op = read\_op$}
 	 \State $rAck \gets rAck \cup \{(s,m)\}$
\EndIf
\newline
\newline
\State {At each writer $w$}
\State \textbf{Variables:}
\State $tg \in \tup{\mathbb{N}^{+},\idSet}$, $v \in V$, $write\_op \in \mathbb{N}^{+}$
\State $maxTS \in \mathbb{N}^{+}$, $mAck \subseteq \srvSet \times M$
\State \textbf{Initialization:}
\State $tg \leftarrow \tup{0,w}$, $v \leftarrow \perp$, $write\_op \leftarrow 0$
\State $maxTS \leftarrow 0$
\Function{Write}{$val: input$}
	\State $write\_op \leftarrow write\_op + 1$		\label{line:omam:srvr:inc_writeop}\label{line:omam:writer:reset:start}
	\State $mAck \leftarrow \emptyset$					\label{line:omam:writer:reset:end}
	\State \textbf{broadcast}$(\tup{{\sf discover}, write\_op, w})$ to $\srvSet$ \label{line:omam:srvr:discover_max:start} \label{line:omam:srvr:bdcast_discover}
	\State \textbf{wait until} $(|mAck| = {|\srvSet|}/{2} +1)$ \label{line:omam:srvr:write_wait_majority_discover} \label{line:omam:srvr:discover_majority:start} \label{line:omam:srvr:discover_majority:end}
	\State $maxTS \leftarrow max\{m.tg.ts^{\prime}| m \in mAck \}$ \label{line:omam:srvr:discover_max:end} \label{line:omam:srvr:find_max}
	\State $(tag,v) \leftarrow (\tup{maxTS+1,w}, val)$ \label{line:omam:srvr:update_tag}
	\State $write\_op \leftarrow write\_op + 1$		\label{line:omam:srvr:inc_writeop_2}
	\State $mAck \leftarrow \emptyset$
	\State \mbox{\textbf{broadcast}$(\tup{{\sf writeRequest}, \tup{tg, v}, write\_op, w})$ to $\srvSet$}\label{line:omam:srvr:bdcast_write_req}
	\State \textbf{wait until} $(|mAck| = {|\srvSet|}/{2} +1)$ \label{line:omam:srvr:write_wait_majority}
	\State \textbf{return}
\EndFunction
\newline
\State \textbf{Upon receive} $m$ from $s$ \label{line:omam:writer:check_fresh:start}
 	 \If{$m.write\_op = write\_op$}
 	 \State $mAck \gets mAck \cup \{(s,m)\}$ \label{line:omam:writer:check_fresh:end}
\EndIf
\pagebreak
\State {At each server $s_i$}
\State \textbf{Variables:}
\State $tg \in \tup{\mathbb{N}^{+},\idSet}$, $v \in V$, 
\State $write\_ops[1...|\wSet|+1]$ : array of int
\State $operations[1...|\rdSet|+1]$ : array of int
\State $relays[1...|\rdSet|+1]$ : array of int
\State \textbf{Initialization:}
\State $tg \leftarrow \tup{0,s_i}$, $v \leftarrow \perp$
\State $write\_ops[i] \leftarrow 0$ for $i \in \wSet$
\State $operations[i] \leftarrow 0$ for $i \in \rdSet$
\State $relays[i] \leftarrow 0$ for $i \in \rdSet$
\newline
\State \textbf{Upon receive}{$(\tup{{\sf readRequest}, r, read\_op})$}{} 	\label{line:omam:srv:readrcv:start}
	\State \textbf{broadcast} $({\sf readRelay}, \tup{tg,v}, r, read\_op, s_i)$ to $\srvSet$	\label{line:omam:srv:readrcv:end}		
\newline 
\State \textbf{Upon receive}{$\tup{({\sf readRelay},\tup{ tg^{\prime}, v^{\prime}}, r, read\_op, s_i)}$}{}
	\If{ $(tg < tg^{\prime})$}						\label{line:omam:srvr:tsupdate:start}		
		\State $ \tup{tg,v }$ $\leftarrow$ $\tup{ tg^{\prime},v^{\prime}} $.
	\EndIf									\label{line:omam:srvr:tsupdate:end}
	\If{ $(operations[r] < read\_op)$}  				 \label{line:omam:srvr:check_read_op:start}
		\State $ operations[r] \leftarrow read\_op $.
		\State $ relays[r] \leftarrow 0 $.
	\EndIf 									\label{line:omam:srvr:check_read_op:end}
	\If{$(operations[r] = read\_op)$} 				\label{line:omam:srvr:inc_relays:start}
		\State $relays[r] \leftarrow relays[r] + 1$ .
	\EndIf									\label{line:omam:srvr:inc_relays:end}
	\If{$(relays[r] = {|\srvSet|}/{2} +1)$}			\label{line:omam:srvr:rdreachmajority:start}	
		\State \textbf{Send} $({\sf readAck}, \tup{tg,v }, read\_op, s_i)$ to $r$
	\EndIf     							\label{line:omam:srvr:rdreachmajority:end}
\newline   
\State \textbf{Upon receive}{$(\tup{{\sf discover}, write\_op, w})$}{} \label{line:omam:srvr:bdcast_discover:start}		
	\State \textbf{Send} $(\tup{{\sf discoverAck}, \tup{tg,v}, write\_op, s_i})$ to $w$ \label{line:omam:srvr:bdcast_discover:end}
\newline
\State \textbf{Upon receive}{$(\tup{{\sf writeRequest}, tg^{\prime}, v^{\prime}, write\_op, w})$}{}
	\If{$((tg < tg^{\prime}) \land (write\_op[w] < write\_op))$}	\label{line:omam:srvr:wrtsupdate:start}	
		\State $ \tup{tg,v}$ $\leftarrow$ $\tup{tg^{\prime}, v^{\prime}}$
		\State $write\_ops[w] \leftarrow write\_op$
	\EndIf														\label{line:omam:srvr:wrtsupdate:end}
	\State \textbf{send} $(\tup{{\sf writeAck}, \tup{tg,v }, write\_op, s_i})$ to  $w$ \label{line:omam:srvr:write_ack}
\end{algorithmic}
}
\vspace*{-\bigskipamount}
\end{multicols}\vspace*{-\bigskipamount}
\end{algorithm}

\noindent{\bf Writer Protocol.} 
This protocol is similar to \cite{LS97}.
When a write operation is invoked, a writer $w$ 
monotonically increases its local write operation counter $write\_op$, empties the set $mAck$ that holds the received acknowledgment messages 
(L\ref{line:omam:writer:reset:start} - \ref{line:omam:writer:reset:end}), 
and it broadcasts a \act{discover} message to all servers $s \in \srvSet$ (L\ref{line:omam:srvr:bdcast_discover}). 
It then waits to collect 
\act{discoverAck} messages from a majority of servers, ${|\srvSet|}/{2} +1$.
While collecting \act{discoverAck} messages, writer $w$ checks the $write\_op$ variable that is included in the message $m$ and discards any message where the value of $write\_op < m.write\_op$ 
(L\ref{line:omam:writer:check_fresh:start} - \ref{line:omam:writer:check_fresh:end}). 
This, happens in order to avoid any delayed \act{discoverAck} messages sent during previous write operations.
Once the {\act{discoverAck}} messages are collected, writer $w$ determines the 
maximum timestamp $maxTS$
from the tags included in the received messages (L\ref{line:omam:srvr:find_max}) and creates a new local tag $\tg{}$,
in which it assigns its id and sets the timestamp to one higher than the maximum one, 
$tg = \tup{maxTS +1, w}$ (L\ref{line:omam:srvr:update_tag}). 
The writer then broadcasts a {\act{writeRequest}} message, including the updated tag, the value to be written, 
its write operation counter and id, $\tg{}$, $v$, $write\_op$ and $w$, to all the participating servers 
(L\ref{line:omam:srvr:bdcast_write_req}). 
It then waits to collect ${|\srvSet|}/{2} +1$ {\act{writeAck}} messages (L\ref{line:omam:srvr:write_wait_majority}) for completion. 



\noindent{\bf Server Protocol.} 
Servers react to messages from the readers exactly as in Algorithm \osam{}.
Here we describe server actions for {\sf{discover}} and {\sf{writeRequest}} messages.
%

$(1)$ Upon receiving message $\tup{{\sf discover},write\_op, w}$, server $\srvr$ attaches its local tag and local value in a new {\act{discoverAck}} message that it sends back to the requesting writer $w$ (L\ref{line:omam:srvr:bdcast_discover:start}-\ref{line:omam:srvr:bdcast_discover:end}).

$(2)$ Upon receiving 
$\tup{{\sf writeRequest} ,\tup{tg^{\prime},v^{\prime}},write\_op, w}$ message 
server compares its local tag $\tg{}$ with the received tag $\tg{}^{\prime}$. 
If the message is not stale 
and server's local tag is older, $\tg{} < \tg{}^{\prime}$, 
it updates its local timestamp and local value to those received
(L\ref{line:omam:srvr:wrtsupdate:start}-\ref{line:omam:srvr:wrtsupdate:end}). 
Otherwise, no update takes place. 
Server $\srvr$ acknowledges the requesting writer $w$ by creating and sending it a {\act{writeAck}} message
(L\ref{line:omam:srvr:write_ack}). 
\subsection{Correctness.}
\label{sec:omam:correct}

To show correctness of Algorithm \ref{alg:omam}
we prove that it satisfies the \emph{termination} and \emph{atomicity} properties.

\noindent{\bf Liveness.} 
Similarly to \ohsam{}, termination holds
with respect to our failure model:
up to $f$ servers may fail, where $f < |\srvSet|/2$
and each operation waits for messages from some majority of servers.
We now give additional details.
\smallskip

\smallskip 

\noindent{\it Write Operation.} 
Writer $w$ finds the maximum tag by broadcasting a {\act{discover}} message to all servers forming exchange \ex{1},
and waiting to collect {\act{discoverAck}} replies from a majority of servers during exchange \ex{2}
(L\ref{line:omam:srvr:discover_max:start}-\ref{line:omam:srvr:discover_max:end} and 
L\ref{line:omam:srvr:bdcast_discover:start}-\ref{line:omam:srvr:bdcast_discover:end}). 
Since we tolerate $f < \frac{|S|}{2}$ crashes, 
then at least a majority of live servers will collect the {\act{discover}} messages from \ex{1} and reply to writer $w$ in \ex{2}.
Once the maximum timestamp is determined, then writer $w$ updates its local tag 
and broadcasts a {\act{writeRequest}} message to all the servers forming \ex{3}.
Writer $w$ then waits to collect {\act{writeAck}} replies from a majority of servers before completion.
Again, at least a majority of servers collects the {\act{writeRequest}} message during \ex{3},
and acknowledges to the requesting writer $w$ in \ex{4}.
\smallskip

\noindent{\it Read Operation.} 
A read operation of algorithm \omam{} differs from \osam{} only by using tags instead of timestamps
in order to impose an ordering on the values written.
The structure of the read protocol is identical to \osam{}, thus 
\emph{liveness} is ensured as reasoned in section \ref{sec:osam:correct}.

Based on the above, any read or write operation collects a 
{sufficient} number of messages to terminate, guaranteeing \emph{liveness}.
 
\smallskip

\paragraph{Atomicity.} 
{\small MWMR} setting we use tags instead of timestamps, and here we show how 
algorithm \omam{} (algorithm \ref{alg:omam}) satisfies \emph{atomicity}  
using \emph{tags}.
We now state and prove the following lemmas.

It is easy to see that the $\tg{}$ variable in each server $\srvr$ is monotonically increasing. 
This leads to the following lemma.

\begin{lemma} 
\label{lem:omam:server_monotonicity}
In any execution $\EX$ of \omam{}, the variable $\tg{}$ maintained by any server $\srvr$ in the system is non-negative and monotonically increasing.
\end{lemma}

\begin{proof}
When server $\srvr$ receives a tag $\tg{}$ then $\srvr$ updates its 
local tag $\tg{s}$ iff $\tg{} > \tg{s}$ (L\ref{line:omam:srvr:tsupdate:start}-\ref{line:omam:srvr:tsupdate:end}
and L\ref{line:omam:srvr:wrtsupdate:start}-\ref{line:omam:srvr:wrtsupdate:end}). 
Thus the local tag of the server monotonically increases
and the lemma follows. 
\end{proof} 
 
 Next, we show that if a read operation $\rd_2$ succeeds read operation $\rd_1$, 
then $\rd_2$ always returns a value at least as recent as the one returned by $\rd_1$.

\begin{lemma}
\label{lem:omam:read-after-read}
In any execution $\EX$ of \omam{}, If $\rd_1$ and $\rd_2$ 
are two read operations such that $\rd_1$ precedes $\rd_2$, i.e., 
$\rd_1 \to \rd_2$, and $\rd_1$ returns a tag $\tg{1}$, then $\rd_2$ returns a tag $\tg{2} \geq \tg{1}$.
\end{lemma}

\begin{proof}
Let the  operations $\rd_1$ and $\rd_2$ be invoked by processes 
$\rdr_1$ and $\rdr_2$ respectively (not necessarily different). 
Let $RSet_1$ and 
$RSet_2$ be the sets of servers that reply to $\rdr_1$ and $\rdr_2$ during $\rd_1$ 
and $\rd_2$ respectively.

Suppose, for purposes of contradiction, that read operations $\rd_1$ and $\rd_2$ exist such that $\rd_2$ succeeds $\rd_1$, i.e., $\rd_1 \to \rd_2$, and the operation $\rd_2$ returns a tag $\tg{2}$ which is smaller than the $\tg{1}$ returned by $\rd_1$, i.e., $\tg{2} < \tg{1}$. 

According to our algorithm,  
$\rd_2$ returns a tag $\tg{2}$ which is smaller than 
the minimum tag received by $\rd_1$, i.e., $\tg{1}$, if 
$\rd_2$ discovers tag $\tg{2}$ and value $v$ in the {\act{readAck}} message of some server 
$\srvr_x\in RSet_2$, and $\tg{2}$ is the minimum tag received by $\rd_2$.

Assume that server $\srvr_x$ replies 
with tag $\tg{}^{\prime}$ and value $v^{\prime}$ 
to read operation $\rd_1$, i.e., $\srvr_x\in RSet_1$. 
By monotonicity of the timestamp 
at the servers (Lemma \ref{lem:omam:server_monotonicity}), and since $\rd_1\bef\rd_2$, then 
it must be the case that $\tg{}^{\prime}\leq \tg{2}$.
According to our assumption $\tg{1}>\tg{2}$, and since $\tg{1}$ is the 
smallest tag sent to $\rd_1$ by any server in $RSet_1$, then it follows that $\rdr_1$ does
not receive the {\act{readAck}} message from $\srvr_x$, and hence $\srvr_x\notin RSet_1$. 

Now examine the actions of  server $\srvr_x$. From the algorithm, 
server $\srvr_x$ collects {\act{readRelay}} messages from a majority 
of servers in $\srvSet$ before sending {\act{readAck}} message to $\rd_2$
(L\ref{line:omam:srvr:rdreachmajority:start}-\ref{line:omam:srvr:rdreachmajority:end}).	
Let $RRSet_{\srvr_x}$ denote the set of servers that send {\act{readRelay}} to $\srvr_x$.
Since, both $RRSet_{\srvr_x}$ and $RSet_1$ contain some majority of the servers 
then it follows that
$RRSet_{\srvr_x}\cap RSet_1\neq\emptyset$.

This above means that there exists a server $\srvr_i\in RRSet_{\srvr_x}\cap RSet_1$
that sends 
(i) a {\act{readAck}} message to $\rdr_1$ for $\rd_1$, and (ii) a {\act{readRelay}} message to 
$\srvr_x$ during $\rd_2$. 
Note that $\srvr_i$ sends a {\act{readRelay}} message for $\rd_2$ only after
it receives a read request from $\rd_2$ 
(L\ref{line:omam:srv:readrcv:start}-\ref{line:omam:srv:readrcv:end}).
Since $\rd_1\bef\rd_2$, then it follows that $\srvr_i$ sends the 
{\act{readAck}} message to $\rd_1$ before sending the {\act{readRelay}} message to $\srvr_x$. Thus,
by Lemma \ref{lem:omam:server_monotonicity}, if $\srvr_i$ attaches a tag $\tg{s_i}$ 
in the {\act{readAck}} to $\rd_1$, then $\srvr_i$ attaches a tag $\tg{\srvr_i}^{\prime}$ 
in the {\act{readRelay}} message to $\srvr_x$, such that  $\tg{\srvr_i}^{\prime}\geq\tg{\srvr_i}$. 
Since $\tg{1}$ is the minimum tag received by $\rd_1$, then $\tg{\srvr_i}\geq \tg{1}$,
then $\tg{\srvr_i}^{\prime}\geq \tg{1}$ as well. 
By Lemma \ref{lem:omam:server_monotonicity}, and since 
$\srvr_x$ receives the {\act{readRelay}} message from $\srvr_i$ before sending a {\act{readAck}}
to $\rd_2$, it follows that $\srvr_x$ sends a tag $\tg{2}\geq\tg{\srvr_i}^{\prime}$.
Therefore, $\tg{2}\geq\tg{1}$ and this contradicts our initial assumption and completes
our proof.
\end{proof}

Next, we reason that if a write operation $\wrt_2$ succeeds write operation $\wrt_1$, 
then $\wrt_2$ writes a value accosiated with a tag strictly higher than $\wrt_1$.

 \begin{lemma}
\label{lem:omam:write-after-write}
In any execution $\EX$ of \omam{}, 
if a write operation $\omega_1$ writes a value with tag $\tg{1}$ then for any succeeding write operation 
$\omega_2$ that writes a value with tag $\tg{2}$ 
we have $\tg{2} > \tg{1}$. 
\end{lemma}

\begin{proof}
Let $WSet_1$ be the set of servers that send a {\act{writeAck}} message 
within write operation $\omega_1$. Let $Disc_2$ be the set of servers that 
send a {\act{discoverAck}} message within  write operation $\omega_2$.

Based on the assumption, write operation $\omega_1$ is complete. 
By Lemma \ref{lem:omam:server_monotonicity}, we know that if a server $\srvr$ receives a tag $\tg{}$ from a process $\pr$ ,
then $\srvr$  includes tag $\tg{}^{\prime}$ s.t. $\tg{}^{\prime} \geq \tg{}$ in any subsequently message.
Thus the servers in $WSet_1$ send a {\act{writeAck}} message within $\omega_1$ 
with tag at least tag $\tg{1}$. 
Hence, every server $\srvr_x \in WSet$ obtains tag $\tg{\srvr_x} \geq \tg{1}$. 

When  write operation $\omega_2$ is invoked, 
it obtains the maximum tag, $max\_tag$, from the tags stored in at least a majority of servers. 
This is achieved by sending {\act{discover}} messages to all servers and 
collecting  {\act{discoverAck}} replies from a majority of servers forming set $Disc_2$ 
(L\ref{line:omam:srvr:discover_max:start}-\ref{line:omam:srvr:discover_max:end} 
and L\ref{line:omam:srvr:bdcast_discover:start}-\ref{line:omam:srvr:bdcast_discover:end}).

Sets $WSet_1$ and $Disc_2$ contain a majority of servers, 
and so $WSet_1\cap Disc_2\neq\emptyset$. 
Thus, by Lemma \ref{lem:omam:server_monotonicity}, any server $\srvr_k \in WSet\cap Disc_2$ has a tag $\tg{\srvr_k}$ s.t. $\tg{\srvr_k} \geq \tg{\srvr_x} \geq \tg{1}$. 
Furthermore, the invoker of $\omega_2$ discovers a $max\_tag$ s.t. 
$max\_tag \geq \tg{\srvr_k} \geq \tg{\srvr_x} \geq \tg{1}$.
The invoker updates its local tag by increasing the maximum tag it discovered,
i.e. $\tg{2} = \tup{max\_tag+1, v}$ (L\ref{line:omam:srvr:update_tag}), 
and associating $\tg{2}$ with the value to be written. 
We know that, $\tg{2} > max\_tag \geq \tg{1}$, hence $local\_tag > \tg{1}$. 

Now the invoker of $\omega_2$ includes its  tag 
$local\_tag$ with {\act{writeRequest}} message
to all servers, and terminates upon receiving {\act{writeAck}} messages 
from a majority of servers. 
By Lemma \ref{lem:omam:server_monotonicity}, $\omega_2$  receives {\act{writeAck}}
messages with a tag $\tg{2}$ s.t. $\tg{2} \geq local\_tag > \tg{1}$ hence $\tg{2} > \tg{1}$, 
and the lemma follows. 
\end{proof}

 At this point we have to show that any read operation 
which succeeds a write operation, will receive {\act{readAck}} 
messages from the servers where each included timestamp 
will be greater or equal to the one that the complete write operation returned. 

\begin{lemma}
\label{lem:omam:read-after-write-step1}
In any execution $\EX$ of \omam{}, if a read operation $\rd$ succeeds a write 
operation $\omega$ that writes value $v$ with tag $\tg{}$, i.e., $\omega \bef \rd$, and 
receives {\act{readAck}} messages from a majority of servers $RSet$, then 
each $\srvr\in RSet$ sends a {\act{readAck}} message to $\rd$ with a tag $\tg{s}$ s.t. $\tg{s} \geq \tg{}$. 
\end{lemma}

\begin{proof}
Let $WSet$ be the set of servers that send a {\act{writeAck}} message to the 
write operation $\omega$ and let $RRSet$ be the set of servers that sent {\act{readRelay}} messages to server $\srvr$.  

It is given that write operation $\omega$ is complete. 
By Lemma \ref{lem:omam:server_monotonicity}, we know that if server $\srvr$ receives a tag $\tg{}$ from process $\pr$,
then $\srvr$ includes a tag $\tg{}^{\prime}$ s.t. $\tg{}^{\prime} \geq \tg{}$ in any subsequent message.
Thus a majority  of servers, forming $WSet$, send a {\act{writeAck}} message 
in $\omega$ with tag greater or equal to tag $\tg{}$. 
Hence, every server $\srvr_x \in WSet$ has a tag $\tg{\srvr_x} \geq \tg{}$.
 Let us now examine tag $\tg{s}$ that server $\srvr$ sends to read operation $\rd$.  

Before server $\srvr$ sends a {\act{readAck}} message to $\rd$,
 it must receive {\act{readRelay}} messages for the majority of servers, $RRSet$ 
(L\ref{line:omam:srvr:rdreachmajority:start}-\ref{line:omam:srvr:rdreachmajority:end}). 
Since both $WSet$ and $RRSet$ contain a majority of servers, then it follows that $WSet\cap RRSet\neq\emptyset$. Thus, by Lemma \ref{lem:omam:server_monotonicity}, any server $\srvr_x \in WSet\cap RRSet$ has a tag $\tg{x}$ s.t. $\tg{x} \geq \tg{}$.

Since server $\srvr_x \in RRSet$
and by the algorithm, server's $\srvr$ tag is always updated to the highest tag it observes 
(L\ref{line:omam:srvr:tsupdate:start}-\ref{line:omam:srvr:tsupdate:end}), 
then when server $\srvr$ receives the message from $\srvr_x$, it  updates its tag $\tg{s}$ s.t. $\tg{s} \geq \tg{x}$.
 
Furthermore, server $\srvr$ creates a {\act{readAck}} message where it includes its local tag $\tg{s}$ 
and its local value $v_s$, and sends this {\act{readAck}} message within the read operation $\rd$ 
(L\ref{line:omam:srvr:rdreachmajority:start}-\ref{line:omam:srvr:rdreachmajority:end}). 
Each $\srvr\in RSet$ sends a {\act{readAck}} to $\rd$ with a tag $\tg{s}$ s.t. $\tg{s} \geq \tg{x} \geq \tg{}$. 
Therefore,  $\tg{s} \geq \tg{}$  and the lemma follows.
\end{proof}

Next we show that if a read operation succeeds a write operation, then it returns a value at least as recent as the one written.

\begin{lemma}
\label{lem:omam:read-after-write}
In any execution $\EX$ of \omam{}, if  read operation $\rd$ succeeds write operation $\omega$,
i.e., $\omega \to \rd$, 
that writes value $v$ associated with tag $\tg{}$,
 and returns tag
 $\tg{}^{\prime}$, then $\tg{}^{\prime} \geq \tg{}$.      
\end{lemma}

\begin{proof}
Suppose that read operation $\rd$ receives {\act{readAck}} messages from a majority of servers $RSet$ 
and 
decides on a tag $\tg{}^{\prime}$ associated with value $v$ and terminates. 

In this case, by Algorithm (L\ref{line:omam:client:reachmajority:start}-\ref{line:omam:client:reachmajority:end})
it follows that read operation $\rd$  decides on a tag $\tg{}^{\prime}$ that belongs to a {\act{readAck}} message among the messages from servers in $RSet$; and it is the minimum tag among all the tags that are included in messages of  servers  $RSet$, hence $\tg{}^{\prime}=min\_tag$. 

Furthermore, since $\tg{}^{\prime} = min\_tag$ holds and from Lemma \ref{lem:omam:read-after-write-step1},  $min\_tag \geq \tg{}$ holds, where $\tg{}$ is the tag returned from the last complete write operation $\omega$, then $\tg{}^{\prime} = min\_tag \geq \tg{}$ also holds. Therefore, $\tg{}^{\prime} \geq \tg{}$ holds and the lemma follows.
\end{proof}

\begin{lemma} 
\label{lem:omam:write-after-read}
In any execution $\EX$ of \omam{}, if a write $\wrt$ succeeds a read operation $\rd$ 
that reads tag $\tg{}$, i.e. $\rd \to \wrt$, and returns a tag $\tg{}^{\prime}$, then $\tg{}^{\prime} > \tg{}$.      
\end{lemma}

\begin{proof}
Let $RR$ be the set of servers that sent {\act{readRelay}} messages to $\rho$.
Let $dAck$ be the set of servers that sent {\act{discoverAck}} messages to $\wrt$.
Let $wAck$ be the set of servers that sent {\act{writeAck}} messages to $\wrt$
and let $RA$ be the set of servers that sent {\act{readAck}} messages to $\rd$. 
It is not necessary that $a \neq b \neq c$ holds.

Based on the read protocol, 
the read operation $\rd$ terminates when it receives
{\act{readAck}} messages from a majority of servers.
It follows that $\rd$ decides on the minimum tag, $\tg{}=minTG$, 
among all the tags in the {\act{readAck}} messages of the set $RA$
and terminates.
Writer $\wrt$, initially it broadcasts a {\act{discover}} message to all servers, and
it then awaits for ``fresh'' {\act{discoverAck}} messages from amajority of servers, that is, set $dAck$.
Each of $RA$ and $dAck$ sets are from majorities of servers,
and so $RA\cap dAck\neq\emptyset$. 
By Lemma \ref{lem:omam:server_monotonicity}, any server $\srvr_k \in RA\cap dAck$ has a tag 
$\tg{\srvr_k}$ s.t. $\tg{\srvr_k} \geq \tg{}$. 
Since $\wrt$ generates a new local tag-value $(\tg{}^{\prime},v)$ pair
in which it assigns the timestamp to be one higher than the one discovered in the \emph{maximum} tag
from set $dAck$,
it follows that $\tg{}^{\prime} > \tg{}$.  
Furthermore, $\wrt$ broadcasts the value to be written associated with $\tg{}^{\prime} $ 
in a {\act{writeRequest}} message to all servers 
and it awaits for {\act{writeAck}} messages from a majority of servers before completion, set $wAck$.
Observe that each of $dAck$ and $wAck$ sets are from majority of servers,
and so $dAck\cap wAck\neq\emptyset$. 
By Lemma \ref{lem:omam:server_monotonicity}, any server $\srvr_k \in dAck\cap wAck$ has a tag 
$\tg{\srvr_k}$ s.t. $\tg{\srvr_k} \geq \tg{}^{\prime} > \tg{} $ and the result follows.
\end{proof}

\begin{theorem}
Algorithm {\ohmam{}} implements an atomic \mw{} object.
\end{theorem}

\begin{proof} 
We use the above lemmas and the operations order definition 
to reason about each of the three \emph{atomicity} conditions A1, A2 and A3. 

\smallskip
\noindent{\bf A1~} For any $\op_1,\op_2\in\Pi$ such that $\op_1\bef\op_2$, it cannot be that $\op_2\prec\op_1$.

\noindent	If both $\op_1$ and $\op_2$ are writes and $\op_1\bef\op_2$ holds,
				then from Lemma~\ref{lem:omam:write-after-write}
				it follows that 
				$\tg{\op_2} > \tg{\op_1}$.
				By the ordering definition $\op_1\prec\op_2$ is satisfied.
				When $\op_1$ is a write, $\op_2$ a read and $\op_1\bef\op_2$ holds,
				then from Lemmas~
				\ref{lem:omam:read-after-write-step1} and 
				\ref{lem:omam:read-after-write}
				it follows that 
				$\tg{\op_2} \geq \tg{\op_1}$.
				By definition $\op_1\prec\op_2$ is satisfied.
				If $\op_1$ is a read, $\op_2$ a write and $\op_1\bef\op_2$ holds,
				then from Lemma~\ref{lem:omam:write-after-read}
				it follows that $\op_2$ always returns a tag $\tg{\op_2}$ s.t. $\tg{\op_2} > \tg{\op_1}$.
				By the ordering definition $\op_1\prec\op_2$ is satisfied.		
				If both $\op_1$ and $\op_2$ are reads and $\op_1\bef\op_2$ holds, then	
				from Lemma~
				\ref{lem:omam:read-after-read}
				it follows that 
				the tag returned from $\op_2$ is always greater or equal to the one returned from $\op_1$.
				$\tg{\op_2} \geq \tg{\op_1}$.
				If $\tg{\op_2} > \tg{\op_1}$, then by the ordering definition $\op_1\prec\op_2$ holds.
				When $\tg{\op_2} = \tg{\op_1}$ then the ordering is not defined but it cannot be that
				$\op_2\prec\op_1$.
				
\smallskip
\noindent{\bf A2~} For any write $\wrt\in\Pi$ and any operation $\op\in\Pi$, then either $\wrt\prec\op$ or $\op\prec\wrt$.
			
\noindent 	If $\tg{\wrt} > \tg{\op}$, then $\op\prec\wrt$ follows directly.
				Similarly, if $\tg{\wrt} < \tg{\op}$ holds, then it follows that $\wrt\prec\op$.
				When $ts_{\wrt} = ts_{\pi}$ holds, then the uniqueness of each tag
				that a writer creates ensures that $\op$ is a read.
				In particular, remember that each tag is a $\tup{ts,id}$ pair, 
				where $ts$ is a natural number and $id$ a writer identifier.
				Tags are ordered lexicographically, first with respect to the timestamp
				and then with respect to the writer id. Since the writer ids are unique, 
				then even if two writes  use the same timestamp $ts$ in the 
				tag pairs they generate, the two tags cannot be equal as they will 
				differ on the writer id. Furthermore, if the two tags are generated 
				by the same writer, then by well-formedness it must be the case 
				that the latter write will contain a timestamp
				larger than any timestap used by that writer before.
				Since $\op$ is a read operation that receives {\sf{readAck}} messages from a majority of servers,
				then the intersection properties of majorities ensure that $\wrt\prec\op$.

\smallskip
\noindent{\bf A3~} Every read operation returns the value of the last write preceding it according to $\prec$ (or the initial value if there is no such write).
	
\noindent	Let $\wrt$ be the last write preceding read $\rho$.
				From our definition it follows that $\tg{\rho} \geq \tg{\wrt}$.
		  		If $\tg{\rho} = \tg{\wrt}$, then $\rho$ returned
		  	  	a value written by $\wrt$ in some servers. 
		    		Read $\rho$ waited for \act{readAck} messages from a majority of servers
				and the intersection properties of majorities ensure that $\wrt$ was the last complete write operation.    
		   		 %
		    		%
		    		If $\tg{\rho} > \tg{\wrt}$ holds, it must be the case that there is a write $\wrt'$ 
		    		that wrote $\tg{\rho}$ and by definition it must hold that $\wrt \prec \wrt^{\prime}\prec\rd$.
		    		Thus, $\wrt$ is not the preceding write and this cannot be the case.
		    		Lastly, if $\tg{\rho} = 0$, no preceding writes exist, and $\rho$ returns the initial value. 
\end{proof}

Having shown liveness and atomicity of algorithm \omam{} the result follows.


\subsection{Performance.}
\label{sec:omam:performance}

Briefly, for algorithm \omam{} write operations take {4} communication exchanges and
read operations take {3} exchanges.
The (worst case) message complexity of read operations
is $|\srvSet|^{2} + 2|\srvSet|$
and the (worst case)  message complexity of write operations
is $4|\srvSet|$.
This follows directly from the structure of the algorithm. We now give additional details. 

\noindent{\bf Operation Latency.}
\emph{Write operation latency:} 
According to algorithm \omam{}, writer $w$ broadcasts a {\act{discover}} message
to all the servers during exchange \ex{1}, and, 
awaits for {\act{discoverAck}} messages from a majority of servers during \ex{2}. 
Once the {\act{discoverAck}} messages are received, then writer $w$ broadcasts 
a {\act{writeRequest}} message to all servers in exchange \ex{3}. 
Lastly, it waits for {\act{writeAck}} messages from a majority of servers in \ex{4}. 
No further communication is required and the write operation terminates. 
Thus any write operation consists of \emph{4} communication exchanges.

\emph{Read operation latency:} 
The structure of the read protocol of \omam{} is identical to \osam{}, thus 
a read operation consists of \emph{3} communication exchanges
as reasoned in Section~\ref{sec:osam:performance}.

%

\noindent{\bf Message Complexity.}
Similarly as in Section~\ref{sec:osam:performance}, 
we measure operation message complexity as the worst case number of exchanged messages 
in each read and write operation. 

\emph{Write operation:} 
A write operation in algorithm \omam{}
takes \emph{4} communication exchanges. 
The first and the third exchanges, \ex{1} and \ex{3}, 
occur when a writer sends {\act{discover}} and {\act{writeRequest}} messages respectively to all servers in $\srvSet$. 
The second and fourth exchanges, \ex{2} and \ex{4}, occur when servers in $\srvSet$ send 
{\act{discoverAck}} and {\act{writeAck}} messages back to the writer. 
Thus, in a write operation there are $4|\srvSet|$ messages exchanged.

\emph{Read operation:} 
The structure of the read protocol of \omam{} is identical to \osam{} thus,
as reasoned in Section~\ref{sec:osam:performance}, 
during a read operation, $|\srvSet|^{2} + 2|\srvSet|$ messages are exchanged.

%
%

\section{Reducing the Latency of Read Operations}
\label{sec:oram:extensions}

In this section we revise the protocol implementing read operations
of algorithms \osam{} and \omam{} 
to yield protocols that implement read operations that terminate in either 
\emph{two} or \emph{three} communication exchanges.
The key idea here is to let the reader determine ``quickly'' that a majority of servers holds
the same timestamp (or tag) and return its associated value.
This is achieved by having the servers send relay messages to each other as
well as to the requesting reader.
While the reader collects the relays and the read acknowledgment messages, 
if it observes in the set of the received relay messages that a majority of servers holds the same timestamp
(or tag),
then it safely returns the associated value and the read operation terminates in \emph{two} communication exchanges. 
If that was not the case, then the reader proceeds similarly to algorithm \osam{}
and terminates in \emph{three} communication exchanges.
We name the revised algorithms as \osamex{} and \omamex{}.
%


\subsection{Algorithm \osamex{} for the {\sc swmr} setting}
\label{sec:osamex:algorithm}

The code for \osamex{} 
that presents the revised read protocol is given in Algorithm~\ref{alg:osamex}. 
In addition, for the servers protocol we show only the changes from algorithm \osam{}.
We now give additional details. 


\begin{algorithm}[!ht]
\caption{\small Read Protocol Changes for SWMR algorithm \osamex{}}
\label{alg:osamex}
\begin{multicols}{2}
{\footnotesize\sf
\begin{algorithmic}[1]
\makeatletter\setcounter{ALG@line}{118}\makeatother
\State \textit{At each reader $r$}	
\State \textbf{Variables:}
\State $ts \in \mathbb{N}^{+}$, $minTS \in \mathbb{N}^{+}$, $v \in V$
\State $read\_op \in \mathbb{N}^{+}$, $rRelay, rAck \subseteq \srvSet \times M$
\State \textbf{Initialization:}
\State $ts \leftarrow 0$, $minTS \leftarrow 0$, $v \leftarrow \perp$, $read\_op \leftarrow 0$ 
\Function{Read}{}
	\State $read\_op \leftarrow read\_op + 1$. 					     
	\State $rAck, rRelay \leftarrow (\emptyset, \emptyset)$
	\State \textbf{bdcast} $({\sf readRequest}, r, read\_op)$ to $\srvSet$ 
	\State \textbf{wait until} $(|rAck| = {|\srvSet|}/{2} +1)$ \textbf{OR} \label{line:osamex:client:new_wait:start} \label{line:osamex:client:new_or_wait}
	\State ~ ~ ~ $(\exists Z \subseteq rRelay : (|Z| \geq  {|\srvSet|}/{2} +1) ~ \land$ \label{osamex:line:osamex:fast-condition:start}
	\State ~ ~ ~ $(\forall (m^{\prime},s^{\prime}), (m^{\prime\prime},s^{\prime\prime}) \in Z : m^{\prime}.ts = m^{\prime\prime}.ts ))$ \label{line:osamex:client:new_wait:end} \label{osamex:line:osamex:fast-condition:end}
	\If{$(rAck = {|\srvSet|}/{2} +1)$} 
		\State $minTS \leftarrow min\{m.ts^{\prime}| m \in rAck \}$ 
		\State $v \gets \{m.val$ $|$ $m\in rAck \land m.ts'=minTS \}$
		\State \textbf{return$(v)$}		
	\Else \label{line:osamex:client:new_if:start}
		\State $v \gets \{m.val$ $|$ $m\in rRelay\}$
		\State \textbf{return}($v$) \label{line:osamex:client:new_if:end}
	\EndIf  
\EndFunction
\pagebreak
\newline
\State \textbf{Upon receive} $m$ from $s$ 
	\If{$m.read\_op = read\_op$} 
		\If{$m.type = {\sf readAck}$}
			\State $rAck \gets rAck \cup \{(s,m)\}$
		\Else
			\State $rRelay \gets rRelay \cup \{(s,m)\}$
		\EndIf	
	\EndIf
\newline
\State \textit{At each server $s_i$}
\State \textbf{Upon receive}{$(\tup{{\sf readRequest}, r, read\_op})$}{}\label{line:osamex:serv-new}
	\State ~ ~ \textbf{bdcast}$(\tup{{\sf readRelay}, ts, v, r, read\_op, s_i})$ to  $\srvSet \cup \{r\}$\label{line:osamex:srv:bdcast_all}					
\newline
\end{algorithmic}
}
\end{multicols}\vspace*{-\bigskipamount}
\end{algorithm}

In order to construct algorithm \osamex{} we modify 
readers and servers protocol of algorithm \osam{}.
In particular, we update the read protocol 
to wait for both {\sf{readRelay}} and {\sf{readAck}} messages 
(L\ref{line:osamex:client:new_wait:start}-\ref{line:osamex:client:new_wait:end}).
Next, in order for the servers protocol to support the 
broadcast of a {\sf{readRelay}} message to all the servers and the reader process
we replace lines 
\ref{line:osam:srv:readrcv:start}-\ref{line:osam:srv:readrequestbdcast} of \osam{}
with lines \ref{line:osamex:serv-new}-\ref{line:osamex:srv:bdcast_all} of \osamex{},
as shown in algorithm~\ref{alg:osamex}.
The combination of those changes yells algorithm \osamex{}.

\ppp{\em Revised Protocol for the Server.} 
The server sends a {\sf{readRelay}} message to all the servers \emph{and} 
to the requesting reader
(L\ref{line:osamex:srv:bdcast_all}).

\ppp{\em Revised Protocol for the Reader.} 
Here we let the reader await for either
(i) {\sf{readAck}} messages
or 
(ii) {\sf{readRelay}} messages
from a majority of servers
(L\ref{line:osamex:client:new_wait:start}-\ref{line:osamex:client:new_wait:end}).
Notice that in latter, all the ${|\srvSet|}/{2} +1$ {\sf{readRelay}} messages
must include the same timestamp to ensure that at least a majority of servers
is informed regarding the last complete write operation
(L\ref{osamex:line:osamex:fast-condition:start}-\ref{osamex:line:osamex:fast-condition:end}).
In addition, since at least a majority of servers is informed regarding the last complete operation, 
then it is safe for the reader to return the value $v$ associated with the 
timestamp $ts$
found in the
{\sf{readRelay}} messages collected
from the majority of servers.
Otherwise, when case (i) holds, then the read protocol proceeds as in \osam{}.

\subsubsection{Correctness.}
\label{sec:osamex:correct}
%
\emph{Liveness} and \emph{atomicity} of the
 revised algorithm \osamex{}
is shown similarly to algorithm \osam{} as reasoned in Section~\ref{sec:osam:correct}.

\paragraph{Liveness.} Termination of Algorithm \osamex{} 
is guaranteed with respect to our failure model:
up to $f$ servers may fail, where $f < |\srvSet|/2$,
and any type of operation awaits for messages from a majority of servers before completion.
We now provide additional details. 

\emph{Read Operation.} 
A read operation of algorithm \osamex{} terminates when the client either
(i) collects {\sf{readRelay}} messages from at least a majority of serves and all of them include the same timestamp;
or
(ii) collects {\sf{readAck}} messages from a majority of servers.
When case (i) occurs then the operation terminates immediately (and faster).
Otherwise, case (ii) holds, the read operation proceeds identically to algorithm 
\osam{} and its termination is ensured as reasoned in Section \ref{sec:osam:correct}.

\emph{Write Operation.} 
A write operation of algorithm \osamex{} is identical to that of  algorithm \osam{}, 
thus \emph{liveness} is guaranteed as reasoned in Section~\ref{sec:osam:correct}. 


\paragraph{Atomicity.} 
Next, we show how algorithm \osamex{} satisfies atomicity.
Due to the similarity of the writer and server protocols of algorithm \osamex{} to those in \osam{},
we state the lemmas and we omit some of the proofs. 
Lemma \ref{lem:osamex:srv:monotonic} shows that the timestamp variable 
$ts$ maintained by each server $\srvr$ in the system is monotonically non-decreasing.

\begin{lemma}
\label{lem:osamex:srv:monotonic}
In any execution $\EX$ of \osamex{}, the variable $ts$ maintained by any server $\srvr$ in the system is non-negative and monotonically increasing.
\end{lemma}

\begin{Proof}
Lemma \ref{lem:osam:srv:monotonic} for algorithm \osam{} also applies to algorithm \osamex{}
because the modification in line \ref{line:osamex:srv:bdcast_all} does not affect the update of the timestamp $ts$ at the server protocol.
\end{Proof}

Next, we show that if a read operation $\rd_2$ succeeds read operation $\rd_1$, 
then $\rd_2$ always returns a value at least as recent as the one returned by $\rd_1$.

\begin{lemma}
\label{lem:osamex:read-after-read}
In any execution $\EX$ of \osamex{}, if $\rd_1$ and $\rd_2$ 
are two read operations such that $\rd_1$ precedes $\rd_2$, i.e., 
$\rd_1 \to \rd_2$, and $\rd_1$ returns the value for timestamp $ts_1$, then $\rd_2$ returns 
the value for timestamp $ts_2 \geq ts_1$.
\end{lemma}

\begin{Proof} 
Since read operations in algorithm \osam{} terminate in 3 communication exchanges
then from Lemma \ref{lem:osam:read-after-read} we know that any two 
non-concurrent 3-exchange read operations satisfy this. 
Thus we  have to show that the lemma holds for the cases where
(i) a 2-exchange read operation $\rd_1$ precedes a 2-exchange read operation $\rd_2$;
(ii) a 2-exchange read operation $\rd_1$ precedes a 3-exchange read operation $\rd_2$; and
(iii) a 3-exchange read operation $\rd_1$ precedes a 2-exchange read operation $\rd_2$.
Let the two operations $\rd_1$ and $\rd_2$ be invoked by processes with identifiers
$\rdr_1$ and $\rdr_2$ respectively (not necessarily different).

\smallskip\noindent
{\it Case} (i).  Let $RRSet_1$ and 
$RRSet_2$ be the sets of servers that send a {\sf{readRelay}} message to $\rdr_1$ and 
$\rdr_2$ during $\rd_1$ and $\rd_2$.
Assume by contradiction
that 2-exchange read operations $\rd_1$ and $\rd_2$ exist such that $\rd_2$ succeeds $\rd_1$, 
i.e., $\rd_1 \to \rd_2$, and the operation $\rd_2$ returns a timestamp $ts_2$ 
that is smaller than the $ts_1$ returned by $\rd_1$, i.e., $ts_2 < ts_1$. 
According to our algorithm, $\rd_2$ returns a timestamp $ts_2$ that is smaller than 
the timestamp that $\rd_1$ returned i.e., $ts_1$, if 
$\rd_2$ received $|\srvSet|/2 +1$ {\sf{readRelay}} messages that all included the same timestamp $ts_2$
and $ts_2$ is smaller than $ts_1$, which is included in $|\srvSet|/2 +1$ {\sf{readRelay}} messages 
received by $\rd_1$.
Since, both $RRSet_1$ and $RRSet_2$ contain some majority of  servers 
then it follows that $RRSet_1\cap RRSet_2\neq\emptyset$.
And since by Lemma \ref{lem:osamex:srv:monotonic} the timestamp variable $ts$ maintained by servers
is monotonically increasing, then it is impossible that 
$\rd_2$ received $|\srvSet|/2 +1$ {\sf{readRelay}} messages that all included the same timestamp $ts_2$
and $ts_2 < ts_1$. In particular, since at least a majority of servers have a timestamp at least as $ts_1$ then 
$\rd_2$ can receive only $|\srvSet|/2$ {\sf{readRelay}} messages with a timestamp $ts_2$ s.t. $ts_2< ts_1$.
Therefore, this contradicts our assumption.

\smallskip\noindent
{\it Case} (ii).  
Since $\rd_1$ is a 2-exchange operation, then $\rdr_1$ receives at least $|\srvSet|/2 +1$ {\sf{readRelay}} 
messages that includes the same timestamp $ts_1$. 
Thus after the completion of $\rd_1$ 
at least a majority of servers hold a timestamp at least as $ts_1$.
In addition, we know that the servers relay to each other and wait for {\sf{readRelay}} messages
from a majority of servers before they send a {\sf{readAck}} message to the reader $\rdr_2$. 
By Lemma \ref{lem:osamex:srv:monotonic} the timestamp variable $ts$ maintained by servers
is monotonically increasing then each server $s_i$ that sends a {\sf{readAck}} message to $\rdr_2$ 
must include a timestamp $ts_{s_i}$ s.t. $ts_{s_i} \geq ts_1$.
Therefore, the minimum timestamp $ts_2$ that $\rdr_2$ can observe in each {\sf{readAck}} message
received from $s_i$ must be $ts_2 \geq ts_{s_i} \geq ts_1$. 
Since a 3-exchange read operation decides on the minimum
timestamp observed in the {\sf{readAck}} responses, 
then reader $\rdr_2$ will decide on a timestamp $ts_2$ s.t. $ts_2 \geq ts_1$.

\smallskip\noindent
{\it Case} (iii).  
Since $\rd_1$ is a 3-exchange operation, then $\rdr_1$ receives at least $|\srvSet|/2 +1$ {\sf{readAck}}
messages that include the minimum  timestamp  $ts_1$.
Servers relay to each other before they send a {\sf{readAck}} message to $\rd_1$ and
timestamps in servers are monotone (Lemma \ref{lem:osamex:srv:monotonic}),
thus after the completion of $\rd_1$ at least a majority of servers, $s_i \in RSet$, hold a timestamp
no smaller than $ts_1$. 
%
Let $RRSet$ be the set of servers that send a {\sf{readRelay}} message to $\rdr_2$ during
$\rd_2$. In order for $\rd_2$ to terminate, based on the read protocol
the size of $RRSet$ must be at least 
$|\srvSet|/2+1$.
Let $ts_{s_i}$  be a  timestamp received from a server $s_i \in RRSet$.
Since $RSet\cap RRSet\neq\emptyset$ then the 2-exchange operation $\rd_2$ that succeeds $\rd_1$
can receive at most $|\srvSet|/2$ (minority)
{\sf{readRelay}} messages with a timestamp $ts_{s_i}$ s.t. $ts_{s_i} < ts_1$.
Thus, when $\rd_2$ terminates it must return a timestamp $ts_2$ s.t. $ts_2 \geq ts_1$
and the lemma follows.
\end{Proof}

Now we show that if a read operation succeeds a write operation, 
then it returns a value at least as recent as the one written.

\begin{lemma}
\label{lem:osamex:read-after-write}
In any execution $\EX$ of the algorithm, if a read $\rd$ succeeds a write operation $\omega$ that writes timestamp $ts$, i.e. $\omega \to \rd$, and returns a timestamp $ts'$, then $ts^{\prime} \geq ts$.      
\end{lemma}

\begin{Proof}
From Lemma \ref{lem:osam:read-after-write} for algorithm \osam{} we know that lemma holds 
if a 3-exchange read operation succeeds a write operation. We now 
 show that the same holds in case where the read operation 
 terminates in 2 exchanges. 

Assume by contradiction that a 2-exchange read operation $\rd$ and a write operation $\omega$ exist 
such that $\rd$ succeeds $\omega$, i.e. $\omega \to \rd$, and $\rd$ returns a timestamp $ts'$ 
that is smaller than $ts$ that $\omega$ wrote, $ts^{\prime} < ts$.
From our algorithm, in order for this to happen, $\rd$ receives $|\srvSet|/2+1$ {\sf{readRelay}}
messages that all include the same timestamp $ts'$ and $ts^{\prime} < ts$.
Since $\omega$ is complete it means that at least a majority of servers
hold a timestamp $ts_s$ s.t. $ts_s \geq ts$. 
Since any two majorities have a non empty intersection, 
this contradicts the assumption that $\rd$ received $|\srvSet|/2+1$ {\sf{readRelay}}
messages that all included the same timestamp $ts'$ where $ts^{\prime} < ts$ and the lemma follows. 
\end{Proof}

\begin{theorem}
\label{thrm:RevisedSW}
Algorithm {\small \sc \osamex{}} implements an atomic {\small SWMR} object.
\end{theorem}

\begin{Proof} 
We now use the lemmas stated above and the operations order definition
to reason about each of the three \emph{atomicity} conditions A1, A2 and A3. 

\ppp{A1}  For any $\op_1,\op_2\in\Pi$ such that $\op_1\bef\op_2$, it cannot be that $\op_2\prec\op_1$.
		
\noindent 	When the two operations $\op_1$ and $\op_2$ are reads and $\op_1\bef\op_2$ holds, then	
				from Lemma~\ref{lem:osamex:read-after-read}
				it follows that the timestamp returned from $\op_2$ is always greater or equal to the one returned from $\op_1$,
				$ts_{\op_2} \geq ts_{\op_1}$.
				If $ts_{\op_2} > ts_{\op_1}$ then by the ordering definition $\op_1\prec\op_2$ is satisfied.
				When $ts_{\op_2} = ts_{\op_1}$ then the ordering is not defined, thus it cannot be the case that 
				$\op_2\prec\op_1$.
				If $\op_2$ is a write, the sole writer generates a new timestamp by
				incrementing the largest timestamp in the system. 
				By well-formedness (see Section \ref{sec:oram:model}), 
				any timestamp generated by the writer for any write operation that 
				precedes $\op_2$ must be smaller than $ts_{\op_2}$.
				Since $\op_1\bef\op_2$, then it holds that $ts_{\op_1} < ts_{\op_2} $. Hence, by the ordering
				definition it cannot be the case that $\op_2\prec\op_1$.
				Lastly, when $\op_2$ is a read and $\op_1$ a write and $\op_1\bef\op_2$ holds, 
				then from Lemma~\ref{lem:osamex:read-after-write}
				it follows that $ts_{\op_2} \geq ts_{\op_1}$.
				By the ordering definition, it cannot hold that $\op_2\prec\op_1$ in this case either.

\ppp{A2} For any write $\wrt\in\Pi$ and any operation $\op\in\Pi$, then either $\wrt\prec\op$ or $\op\prec\wrt$.

\noindent	
				If the timestamp returned from $\wrt$ is greater than the one returned from $\op$,
				i.e. $ts_{\wrt} > ts_{\op}$,
				 then $\op\prec\wrt$ follows directly.
				Similartly, if $ts_{\wrt} < ts_{\pi}$ holds, then $\wrt\prec\op$ follows.
				If $ts_{\wrt} = ts_{\pi}$, then it must be that $\op$ is a read 
				and 
				$\op$ either discovered $ts_{\wrt}$ as the \emph{minimum} timestamp in at least a majority of servers
				or returned fast $ts_{\wrt}$ because it was noticed in at least a majority of servers.	
				Thus, $\wrt\prec\op$ follows.
			
\ppp{A3} Every read operation returns the value of the last write preceding it according to $\prec$ (or the initial value if there is no such write).
	
\noindent	Let $\wrt$ be the last write preceding read $\rho$.
				From our definition it follows that $ts_{\rho} \geq ts_{\wrt}$.
		    		If $ts_{\rho} = ts_{\wrt}$, then $\rho$ returned
		   		 the value written by $\wrt$ on a majority of servers.
		    		If $ts_{\rho} > ts_{\wrt}$, then it means that $\rho$ obtained a larger 
		    		timestamp. However, the larger timestamp can only be originating from a write
		    		that succeeds $\wrt$, thus $\wrt$ is not the preceding write and this cannot be the case.
		    		Lastly, if $ts_{\rho} = 0$, no preceding writes exist, and $\rho$ returns the initial value. 

\end{Proof}

Having shown liveness and atomicity of algorithm \osamex{} the result follows.


\subsubsection{Performance.}
\label{sec:osamex:performance}

In algorithm \osamex{} write operations take {2} \metrics{} and
read operations take 2 or {3} \metrics{}.
The (worst case) message complexity of read operations
is $|\srvSet|^{2} + 3|\srvSet|$
and the (worst case)  message complexity of write operations
is $2|\srvSet|$. 
These results follows directly from the structure of the algorithm.
We now provide additional details.

\ppp{Operation Latency.}
\emph{Write operation latency:} 
The structure of the write protocol of \osamex{} is identical to \osam{}, thus 
a write operation consists of \emph{2} communication exchanges
as reasond in Section~\ref{sec:osam:performance}.
%
%

\emph{Read operation latency:} 
A reader sends a {\sf{readRequest}} message to all servers
in the first communication exchange \x{1}.
Once the servers receive the {\sf{readRequest}}
message they broadcast a {\sf{readRelay}} message to 
all the  servers \emph{and}
to the requesting reader in the exchange \x{2}.
The reader can terminate at the end of the second exchange, \x{2},
if it can be ``fast'' and complete.
If not, then the operation waits for exchange \x{3} as in
algorithm \osam{} before completion.
Thus, a read operation terminate either in \emph{2} or \emph{3} communication exchanges.

\ppp{Message Complexity.}
\emph{Write operation:} 
The structure of the write protocol of \osamex{} is identical to \osam{}, thus, 
as reasond in Section~\ref{sec:osam:performance},
 $4|\srvSet|$ messages are exchanged during a write operation. 

\emph{Read operation:} 
Read operations in the worst case take \emph{3} communication exchanges. 
Exchange \x{1} occurs when a reader sends a {\sf{readRequest}} message 
to all servers in $\srvSet$. 
The second exchange \x{2} occurs when servers in $\srvSet$ send {\sf{readRelay}} 
messages to all servers in $\srvSet$ and to the requesting reader. 
The final exchange \x{3} occurs when servers in $\srvSet$ send 
a {\sf{readAck}} message to the reader. 
Summing together $|\srvSet| + (|\srvSet|^{2} + |\srvSet|)+ |\srvSet|$, 
shows that in the worst case, $|\srvSet|^{2} + 3|\srvSet|$ 
messages are exchanged during a read operation.


\subsection{Algorithm \omamex{} for the {\sc mwmr} setting}
\label{sec:omamex:algorithm}

Algorithm \omamex{} is obtained similarly to \osamex{} by 
(i) using tags instead of timestamps in the revised read protol of \osamex{} 
and 
(ii) using the write protocol of \omam{} without any modifications.
Next, we reason about \omamex{} correctness.


\subsubsection{Correctness.}
\label{sec:omamex:correct}

\emph{Liveness} and \emph{atomicity} of the
revised algorithm \omamex{}
is shown similarly to algorithm \omam{} as reasoned in Section~\ref{sec:omam:correct}.

\paragraph{Liveness.} Termination of Algorithm \omamex{} 
is guaranteed with respect to our failure model:
up to $f$ servers may fail, where $f < |\srvSet|/2$,
and operations wait for messages only from a majority of servers.
We now give additional details. 

\emph{Read Operation.} 
A read operation of \omamex{} differs from \osamex{} by using tags instead of timestamps
in order to impose an ordering on the values written.
The structure of the read protocol is identical to \osamex{}, thus 
\emph{liveness} is ensured as reasoned in section \ref{sec:osam:correct}.

\emph{Write Operation.} 
Since the write protocol of algorithm \omamex{} is identical to the one 
that algorithm \omam{} uses, \emph{liveness} 
is guaranteed as discussed in Section~\ref{sec:omam:correct}.

\paragraph{Atomicity.} 
Next, we show how algorithm \omamex{} satisfies atomicity.
Due to the similarity of the writer and server protocols of algorithm \omamex{} to those in \omam{},
we state the lemmas and we omit some of the proofs. 
Lemma \ref{lem:omamex:srv:monotonic} shows that the timestamp variable 
$ts$ maintained by each server $\srvr$ in the system is monotonically non-decreasing.

\begin{lemma} 
\label{lem:omamex:srv:monotonic}
In any execution $\EX$ of \omamex{}, the variable $\tg{}$ maintained by any server $\srvr$ in the system is non-negative and monotonically increasing.
\end{lemma}

\begin{Proof}
Lemma \ref{lem:omam:server_monotonicity} for algorithm \omam{} also applies to algorithm \omamex{}
because the modification in line \ref{line:osamex:srv:bdcast_all} does not affect the update of the tag $\tg{}$ at the server protocol.
\end{Proof}

Next, we show that if a read operation $\rd_2$ succeeds read operation $\rd_1$, 
then $\rd_2$ always returns a value at least as recent as the one returned by $\rd_1$.

\begin{lemma}
\label{lem:omamex:read-after-read}
In any execution $\EX$ of \omamex{}, If $\rd_1$ and $\rd_2$ 
are two read operations such that $\rd_1$ precedes $\rd_2$, i.e., 
$\rd_1 \to \rd_2$, and $\rd_1$ returns a tag $\tg{1}$, then $\rd_2$ returns a tag $\tg{2} \geq \tg{1}$.
\end{lemma}

\begin{Proof}
Since read operations in algorithm \omam{} terminate in 3 communication exchanges
then from Lemma \ref{lem:omam:read-after-read} we know that the any 
two non-concurrent 3-exchange satisfy this.
 Thus we now have to show that the lemma holds for the cases where
(i) a 2-exchange read operation $\rd_1$ precedes a 2-exchange read operation $\rd_2$;
(ii) a 2-exchange  read operation $\rd_1$ precedes a 3-exchange read operation $\rd_2$; and
(iii) a 3-exchange read operation $\rd_1$ precedes a 2-exchange read operation $\rd_2$.
Let the two operations $\rd_1$ and $\rd_2$ be invoked by processes with identifiers
$\rdr_1$ and $\rdr_2$ respectively (not necessarily different).

\smallskip
\noindent
{\it Case} (i). 
Assume by contradiction
that 2-exchange read operations $\rd_1$ and $\rd_2$ exist such that $\rd_2$ succeeds $\rd_1$, 
i.e., $\rd_1 \to \rd_2$, and  operation $\rd_2$ returns a tag $\tg{2}$ 
that is smaller than tag $\tg{1}$ returned by $\rd_1$, i.e., $\tg{2} < \tg{1}$. 
Since both operations $\rd_1$ and $\rd_2$ complete in 2 exchanges,  they both
have to collect $|\srvSet|/2 +1$ {\sf{readRelay}} messages with the same tag 
$\tg{1}$ and $\tg{2}$ respectively. 
We know that after the completion or $\rd_1$ at least $|\srvSet|/2 +1$ servers have
a tag at least as $\tg{1}$.
By monotonicity of the tag at the servers (Lemma \ref{lem:omamex:srv:monotonic})
and the fact that $\rd_1$ is completed it follows that it is impossible for $\rd_2$ to collect
$|\srvSet|/2 +1$ {\sf{readRelay}} messages with the same tag $\tg{2}$
s.t. $\tg{2} < \tg{1}$. 
In particular, $\rd_2$ can receive only $|\srvSet|/2$ {\sf{readRelay}} messages 
with a timestamp $\tg{2}$ s.t. $\tg{2} < \tg{1}$. Therefore, this contradicts our assumption.

\smallskip
\noindent
{\it Case} (ii).
We know that since $\rd_1$ is a 2-exchange operation then $\rdr_1$ receives at least $|\srvSet|/2 +1$ {\sf{readRelay}} 
messages that include the same tag $\tg{1}$. 
Thus after the completion of $\rd_1$ 
at least a majority of servers hold a timestamp at least as $\tg{1}$.
Servers relay to each other and wait for {\sf{readRelay}} messages
from a majority of servers before they send a {\sf{readAck}} message to the reader $\rdr_2$. 
By Lemma \ref{lem:omamex:srv:monotonic} since the tag variable $\tg{s}$ maintained by servers
is monotonically increasing then each server $s_i$ that sends a {\sf{readAck}} message to $\rdr_2$ 
must include a tag $\tg{s_i}$ s.t. $\tg{s_i} \geq \tg{1}$.
Therefore, the minimum tag $\tg{2}$ that $\rdr_2$ can observe in each {\sf{readAck}} message
received from $s_i$ must be $\tg{2} \geq \tg{s_i} \geq \tg{1}$. 
Since a 3-exchange read operation decides on the minimum
tag observed in the {\sf{readAck}} responses, 
 reader $\rdr_2$  decides on a timestamp $\tg{2}$ s.t. $\tg{2} \geq \tg{1}$.
 
\smallskip
\noindent
{\it Case} (iii).  
 Since $\rd_1$ is a 3-exchange operation, $\rdr_1$ receives at least $|\srvSet|/2 +1$ {\sf{readAck}}
messages that include the minimum tag $\tg{1}$.
Servers relay to  each other before they send a {\sf{readAck}} message to $\rd_1$,
then, by the monotonicity of  tags at servers (Lemma \ref{lem:omamex:srv:monotonic}),
after the completion of $\rd_1$ at least a majority of servers $s_i \in RSet$ hold a tag at least as $\tg{1}$.
Let $RRSet$ be the set of servers that send a {\sf{readRelay}} message to $\rdr_2$ during
$\rd_2$. In order for $\rd_2$ to terminate
the size of $RRSet$ must be at least 
$|\srvSet|/2+1$
  based on the read protocol.
Let $\tg{s_i}$ be a tag received from a server $s_i \in RRSet$.
Since $RSet_1\cap RRSet\neq\emptyset$ then the 2-exchange operation $\rd_2$ that succeeds $\rd_1$
can receive at most $|\srvSet|/2$ (minority)
{\sf{readRelay}} messages with a tag $\tg{s_i}$ s.t. $\tg{s_i} < \tg{1}$.
Thus, 
when $\rd_2$ terminates it must return a 
tag $\tg{2}$ s.t. $\tg{2} \geq \tg{1}$ 
and the lemma follows.
\end{Proof}

Now we show that if a read operation succeeds a write operation, then it returns a value at least as recent as the one written.

\begin{lemma}
\label{lem:omamex:read-after-write}
In any execution $\EX$ of \omamex{}, if  read operation $\rd$ succeeds write operation $\omega$ 
(i.e., $\omega \to \rd$)
that writes value $v$ associated with  tag $\tg{}$ and returns  tag $\tg{}^{\prime}$, then $\tg{}^{\prime} \geq \tg{}$.      
\end{lemma}

\begin{Proof}
From Lemma \ref{lem:omam:read-after-write} for algorithm \omam{} we know that the lemma holds 
if a 3-exchange read operation succeeds a write operation. We now 
 show that the same holds for 2-exchange read operations. 

Assume by contradiction that a 2-exchange read operation $\rd$ and a write operation $\omega$ exist 
such that $\rd$ succeeds $\omega$, i.e. $\omega \to \rd$, and $\rd$ returns a tag $\tg{}'$ 
that is smaller than the tag $\tg{}$ that $\omega$ wrote, $\tg{}' < \tg{}$.
From the algorithm, in order for this to happen, $\rd$ receives $|\srvSet|/2+1$ {\sf{readRelay}}
messages that all include the same tag $\tg{}'$ and $\tg{}' < \tg{}$.
Since $\omega$ is complete it means that at least a majority of servers
hold tag $\tg{s}$ s.t. $\tg{s} \geq \tg{}$. 
Since any two majorities  intersect, 
this contradicts the assumption that $\rd$ receives $|\srvSet|/2+1$ {\sf{readRelay}}
messages that all include the same timestamp $\tg{}'$ where $\tg{}' < \tg{}$ and the lemma follows. 
\end{Proof}

Next, we reason that if a write operation $\wrt_2$ succeeds write operation $\wrt_1$, 
then $\wrt_2$ writes a value accosiated with a tag strictly higher than $\wrt_1$.

\begin{lemma}
\label{lem:omamex:write-after-write}
In any execution $\EX$ of \omamex{}, 
if a write operation $\omega_1$ writes a value with tag $\tg{1}$ then for any succeeding write operation 
$\omega_2$ that writes a value with tag $\tg{2}$ 
we have $\tg{2} > \tg{1}$. 
\end{lemma}

\begin{Proof}
The modifications of \omamex{} do not have an impact on the write operations
thus this lemma follows directly from lemma~\ref{lem:omam:write-after-write} of \omam{}.
\end{Proof}

\begin{lemma} 
\label{lem:omamex:write-after-read}
In any execution $\EX$ of \omamex{}, if a write $\wrt$ succeeds a read operation $\rd$ 
that reads tag $\tg{}$, i.e. $\rd \to \wrt$, and returns a tag $\tg{}^{\prime}$, then $\tg{}^{\prime} > \tg{}$.      
\end{lemma}

\begin{Proof}
The case where the read operation takes three communication exchanges to terminate
is identical as in lemma~\ref{lem:omam:write-after-read}
of algorithm \omam{}. Thus, we are interested to examine the case where the read terminates
in two communication exchanges. 

Let $RR$ be the set of servers that sent {\sf{readRelay}} messages to $\rho$.
Let $dAck$ be the set of servers that sent {\sf{discoverAck}} messages to $\wrt$.
Let $wAck$ be the set of servers that sent {\sf{writeAck}} messages to $\wrt$
and let $RA$ be the set of servers that sent {\sf{readAck}} messages to $\rd$. 
It is not necessary that $a \neq b \neq c$ holds.

In the case we examine, the read operation $\rd$ terminates when it receives
{\sf{readRelay}} messages from a majority of servers and 
$\rd$ decides on a tag that all servers attached in the set $RA$
and lastly it terminates.
Writer $\wrt$, initially it broadcasts a {\sf{discover}} message to all servers, and
it then awaits for ``fresh'' {\sf{discoverAck}} messages from amajority of servers, that is, set $dAck$.
Each of $RA$ and $dAck$ sets are from majorities of servers,
and so $RA\cap dAck\neq\emptyset$. 
By Lemma \ref{lem:omamex:srv:monotonic}, any server $\srvr_k \in RA\cap dAck$ has a tag 
$\tg{\srvr_k}$ s.t. $\tg{\srvr_k} \geq \tg{}$. 
Since $\wrt$ generates a new local tag-value $(\tg{}^{\prime},v)$ pair
in which it assigns the timestamp to be one higher than the one discovered in the \emph{maximum} tag
from set $dAck$,
it follows that $\tg{}^{\prime} > \tg{}$.  
Furthermore, $\wrt$ broadcasts the value to be written associated with $\tg{}^{\prime} $ 
in a {\sf{writeRequest}} message to all servers 
and it awaits for {\sf{writeAck}} messages from a majority of servers before completion, set $wAck$.
Observe that each of $dAck$ and $wAck$ sets are from majority of servers,
and so $dAck\cap wAck\neq\emptyset$. 
By Lemma \ref{lem:omam:server_monotonicity}, any server $\srvr_k \in dAck\cap wAck$ has a tag 
$\tg{\srvr_k}$ s.t. $\tg{\srvr_k} \geq \tg{}^{\prime} > \tg{} $ and the result follows.

\end{Proof}

Similarly to Theorem \ref{thrm:RevisedSW} we show the following for
algorithm \omamex{}.

\begin{theorem}
Algorithm {\small \sc \omamex{}} implements an atomic {\small MWMR} object.
\end{theorem}
%
%

\begin{Proof} 
We use the above lemmas and the operations order definition 
(using tags instead of timestamps)
to reason about each of the three \emph{atomicity} conditions A1, A2 and A3. 

\ppp{A1} For any $\op_1,\op_2\in\Pi$ such that $\op_1\bef\op_2$, it cannot be that $\op_2\prec\op_1$.

\noindent	If both $\op_1$ and $\op_2$ are writes and $\op_1\bef\op_2$ holds,
				then from Lemma~\ref{lem:omamex:write-after-read}
				it follows that 
				$\tg{\op_2} > \tg{\op_1}$.
				By the ordering definition $\op_1\prec\op_2$ is satisfied.
				When $\op_1$ is a write, $\op_2$ a read and $\op_1\bef\op_2$ holds,
				then from Lemma~
				\ref{lem:omamex:read-after-write}
				it follows that 
				$\tg{\op_2} \geq \tg{\op_1}$.
				By definition $\op_1\prec\op_2$ is satisfied.
				If $\op_1$ is a read, $\op_2$ a write and $\op_1\bef\op_2$ holds,
				then from Lemma~\ref{lem:omamex:write-after-read}
				it follows that $\op_2$ always returns a tag $\tg{\op_2}$ s.t. $\tg{\op_2} > \tg{\op_1}$.
				By the ordering definition $\op_1\prec\op_2$ is satisfied.		
				If both $\op_1$ and $\op_2$ are reads and $\op_1\bef\op_2$ holds, then	
				from Lemma~
				\ref{lem:omamex:read-after-read}
				it follows that 
				the tag returned from $\op_2$ is always greater or equal to the one returned from $\op_1$.
				$\tg{\op_2} \geq \tg{\op_1}$.
				If $\tg{\op_2} > \tg{\op_1}$, then by the ordering definition $\op_1\prec\op_2$ holds.
				When $\tg{\op_2} = \tg{\op_1}$ then the ordering is not defined but it cannot be that
				$\op_2\prec\op_1$.

\ppp{A2} For any write $\wrt\in\Pi$ and any operation $\op\in\Pi$, then either $\wrt\prec\op$ or $\op\prec\wrt$.
			
\noindent 	If $\tg{\wrt} > \tg{\op}$, then $\op\prec\wrt$ follows directly.
				Similarly, if $\tg{\wrt} < \tg{\op}$ holds, then it follows that $\wrt\prec\op$.
				When $ts_{\wrt} = ts_{\pi}$ holds, then the uniqueness of each tag
				that a writer creates ensures that $\op$ is a read.
				In particular, remember that each tag is a $\tup{ts,id}$ pair, 
				where $ts$ is a natural number and $id$ a writer identifier.
				Tags are ordered lexicographically, first with respect to the timestamp
				and then with respect to the writer id. Since the writer ids are unique, 
				then even if two writes  use the same timestamp $ts$ in the 
				tag pairs they generate, the two tags cannot be equal as they will 
				differ on the writer id. Furthermore, if the two tags are generated 
				by the same writer, then by well-formedness it must be the case 
				that the latter write will contain a timestamp
				larger than any timestap used by that writer before.
				Since $\op$ is a read operation that receives either (i) {\sf{readAck}} messages from a majority of servers,
				or (ii) {\sf{readRelay}} messages from a majority of servers with the same $\tg{}$,
				then the intersection properties of majorities ensure that $\wrt\prec\op$.

\ppp{A3} Every read operation returns the value of the last write preceding it according to $\prec$ (or the initial value if there is no such write).
	
\noindent	Let $\wrt$ be the last write preceding read $\rho$.
				From our definition it follows that $\tg{\rho} \geq \tg{\wrt}$.
		  		If $\tg{\rho} = \tg{\wrt}$, then $\rho$ returned
		  	  	a value written by $\wrt$ in some servers. 
		    		Read $\rho$ waited either (i) for {\sf{readAck}} messages from a majority of servers, or 
		    		(ii) {\sf{readRelay}} messages from a majority of servers with the same $\tg{}$.
				Thus the intersection properties of majorities ensure that $\wrt$ was the last complete write operation.    
		   		 %
		    		%
		    		If $\tg{\rho} > \tg{\wrt}$ holds, it must be the case that there is a write $\wrt'$ 
		    		that wrote $\tg{\rho}$ and by definition it must hold that $\wrt \prec \wrt^{\prime}\prec\rd$.
		    		Thus, $\wrt$ is not the preceding write and this cannot be the case.
		    		Lastly, if $\tg{\rho} = 0$, no preceding writes exist, and $\rho$ returns the initial value. 
\end{Proof}

Having shown liveness and atomicity of algorithm \omamex{} the result follows.


\subsubsection{Performance.}
\label{sec:omamex:performance}

In algorithm \omamex{} write operations take {4} \metrics{} and
read operations take {2} or {3} \metrics{}.
The (worst case) message complexity of read operations
is $|\srvSet|^{2} + 3|\srvSet|$
and the (worst case)  message complexity of write operations
is $4|\srvSet|$.
We now provide additional details.

\ppp{Operation Latency.} 
\emph{Write operation latency:} 
The structure of the write protocol of \omamex{} is identical to \omam{}, thus 
a write operation consists of \emph{4} communication exchanges
as reasond in Section~\ref{sec:omam:performance}.

\emph{Read operation latency:} 
The structure of the read protocol of \omamex{} is identical to \osamex{}, thus 
a read operation consists of \emph{at most} \emph{3} communication exchanges
as reasond in Section~\ref{sec:osamex:performance}.

\ppp{Message Complexity.}
\emph{Write operation:} 
The structure of the write protocol of \omamex{} is identical to \omam{}, thus,
as reasoned in Section~\ref{sec:omam:performance},
$4|\srvSet|$ messages are exchanged in a write operation. 
%

\emph{Read operation:} 
The structure of the read protocol of \omamex{} is identical to \osamex{}, thus,
as reasoned in Section~\ref{sec:osamex:performance},
$|\srvSet|^{2} + 3|\srvSet|$ messages are exchanged during a read operation. 
%

\section{Empirical Evaluations}
\label{sec:oram:simulation}
Here we present a comparative study if our algorithms by simulating
them using the NS3 discrete event simulator \cite{NS3}. 
%
We implemented the following three {\small SWMR} algorithms: 
\ABD{} \cite{ABD96}, \osam{}, and \osamex{}.
We also implemented the corresponding three {\small MWMR} algorithms: 
\ABD{}-MW (following the multi-writer extension \cite{LS97}), \omam{}, and \omamex{}.
For comparison  we also implemented a benchmark, called \benchmark{},
that mimics the minimum message requirements for the  {\small SWMR} and {\small MWMR} settings
but without performing any computation or ensuring consistency.  
In particular, \benchmark{} performs two communication exchanges for read and write operations
thus providing a lower bound on performance in simulated scenarios.
Note that \benchmark{}, does not serve
the properties of Atomicity and its use is strictly serving comparison purposes. 
\begin{wrapfigure}{R}{0.54\textwidth}
\vspace*{-5mm}
\includegraphics[width= 0.54\textwidth]{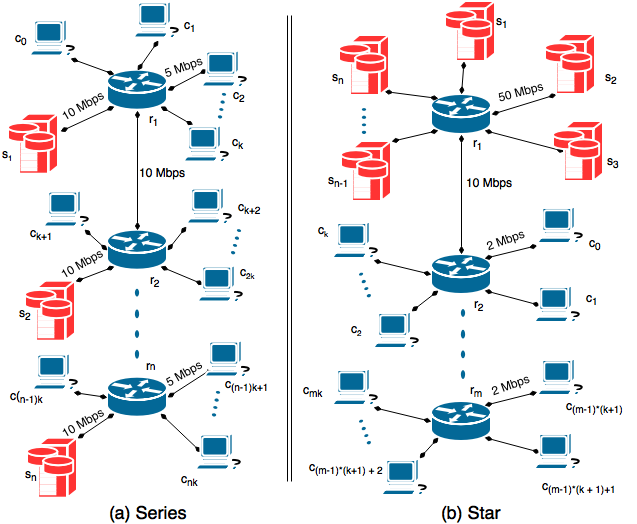}
\caption{Simulated topologies.}
\label{fig:oram:topologies}
\vspace{-5mm}
\end{wrapfigure}

\noindent{\bf Experimentation Setup.}
The experimental configuration consists 
of a single ({\small SWMR}) or multiple ({\small MWMR}) writers, 
a set of readers, and a set of servers.
We assume that at most one server may fail. 
This is done to subject the system to a high communication burden.
Communication among the nodes is established via point-to-point bidirectional links
implemented with a DropTail queue.

For our evaluation,
we use simulations representing two different topologies, 
\emph{Series} and \emph{Star},  that include the same array of routers
but differ in the deployment of server nodes.
In both topologies clients are connected to routers over 5Mbps links
with 2ms delay, the routers are connected over 10Mpbs links with 
4ms delay. 
In the \emph{Series} topology in Figure \ref{fig:oram:topologies}(a),
a server is connected to each router over 10Mbps bandwidth with 2ms delay. 
This topology models a network where servers are separated and appear to be in different networks. 
In the \emph{Star} topology in Figure \ref{fig:oram:topologies}(b)  all  
servers are connected to a single router over 50Mbps links with 2ms delay,
modeling a network where servers are in a close proximity and are well-connected,
e.g., as in a datacenter.
In both topologies readers and writer(s) are located uniformly
with respect to the routers.
We ran NS3 on a Mac OS X with 2.5Ghz Intel Core i7 processor.
The results are compiled as averages over  five samples per each scenario.


\begin{figure}[th]
{\small
\begin{tabular}{c c}
		\includegraphics[width=0.5\textwidth]{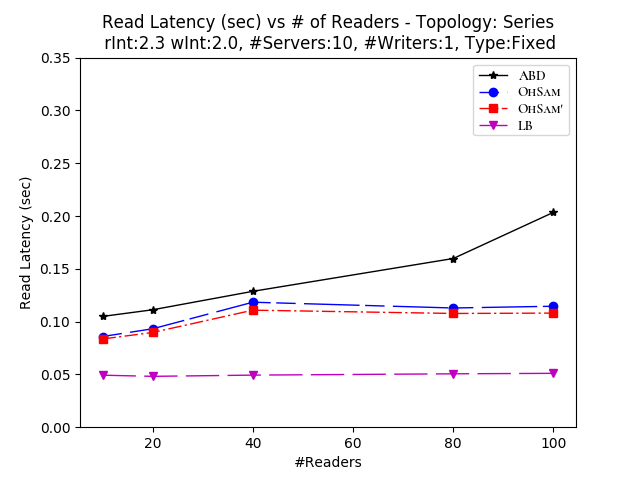}
		&
		\includegraphics[width=0.5\textwidth]{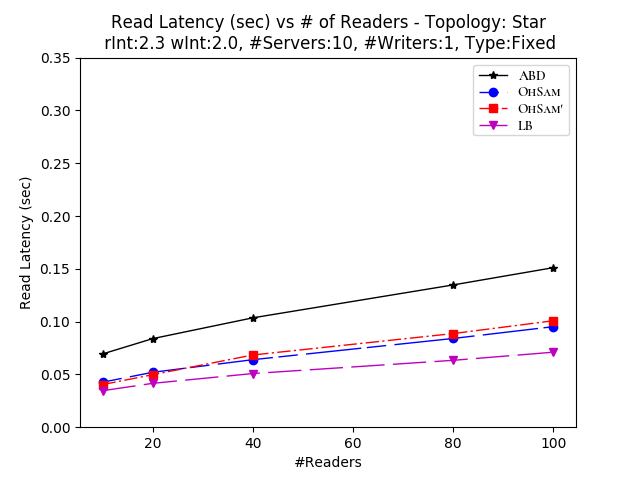} \\
		(a)  & (b) \\
			\includegraphics[width=0.5\textwidth]{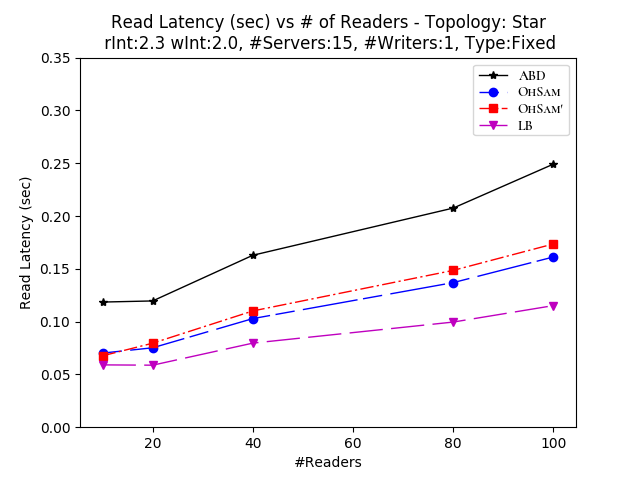} 
		&		 
			\includegraphics[width=0.5\textwidth]{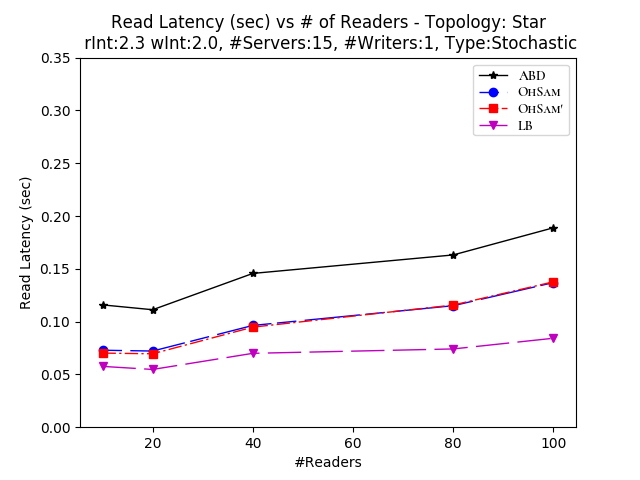}\\
				(c) & (d) \\ 
\end{tabular}
}
\vspace{-\smallskipamount}
\caption{ {\small SWMR} Simulation Results.}
\label{fig:oram:swmrplots}
\vspace{-\smallskipamount}
\end{figure}

\noindent{\bf Performance.}
We assess algorithms in terms of \emph{operation latency} that depends on 
communication delays and local computation delays. 
%
NS3 supports simulated time events, but does not measure delays due to local computation.
In order to measure operation latency we combine two clocks: 
the simulation clock to measure communication delays, and 
a real time clock to measure computation delays.
The sum of the two times yields operation latency.

\noindent{\bf Scenarios.}
To measure performance we define several scenarios.
The scenarios are designed to test
$(i)$ the scalability of the algorithms as the number of readers, writers, and servers increases;
$(ii)$ the contention effect on efficiency, by running different concurrency scenarios; and 
$(iii)$ the effects of chosen topologies on the efficiency.
For scalability we test with the number of readers $|\rdSet|$ from the set $\{10, 20, 40, 80, 100\}$
and the number of servers $|\srvSet|$ from the set $\{10, 15, 20, 25, 30\}$. 
For the {{\small MWMR}} setting
we use at most $80$ readers and we range the number of writers $|\wSet|$
over the set $\{10, 20, 40\}$.
To test contention we set the frequency of each read and write operation to be constant
and we define two different invocation schemes. 
We issue reads  every $rInt=2.3$ seconds and write operations every $wInt=4$ seconds.
We define two invocation schemes:
\emph{fixed} and 
\emph{stochastic}. 
In the \emph{fixed} scheme all  operations are scheduled periodically at a constant interval. 
In the \emph{stochastic} scheme read and write operations are scheduled randomly from
the intervals $[1,rInt]$ and $[1,wInt]$ respectively. 
To test the effects of topology we run our simulations 
using  the \emph{Series} and \emph{Star} topologies.

\noindent{\bf Results.}
We generally observe that
the proposed algorithms
outperform algorithms \ABD{} and \ABDmwmr{} in most scenarios
by a factor of 2. 
A closer examination yields the following observations.


\textit{Scalability:} 
As seen in Figures \ref{fig:oram:swmrplots}(b) and \ref{fig:oram:swmrplots}(c),
increasing the number of readers and servers 
increases latency in the  {\small SWMR} algorithms. 
%
The same observation holds for the {\small MWMR} algorithms.
When the number of the participating readers and writers is reduced then
not surprisingly the latency improves, but the relative performance of the
algorithms remains the same.

%

\begin{figure}[th]
{\small
\begin{tabular}{c c}
		\includegraphics[width=0.5\textwidth]{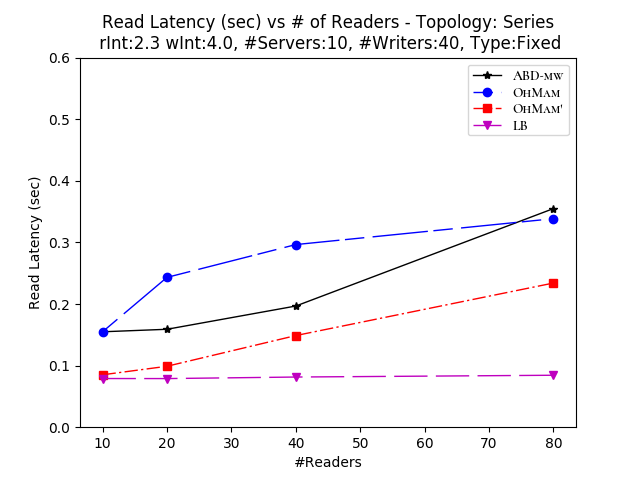}
		&
		\includegraphics[width=0.5\textwidth]{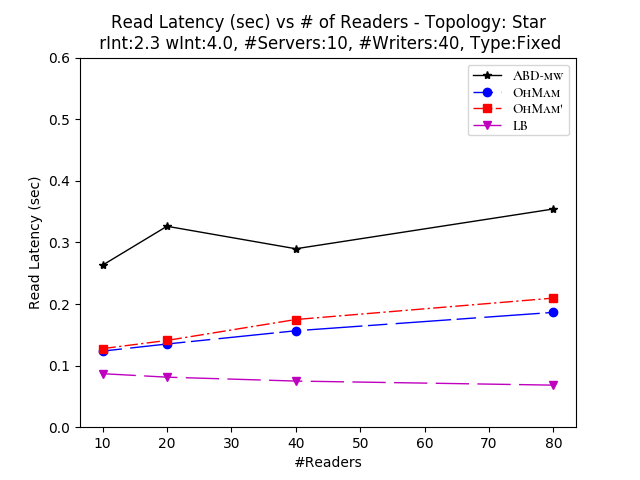} \\
		(e) & (f) \\
			\includegraphics[width=0.5\textwidth]{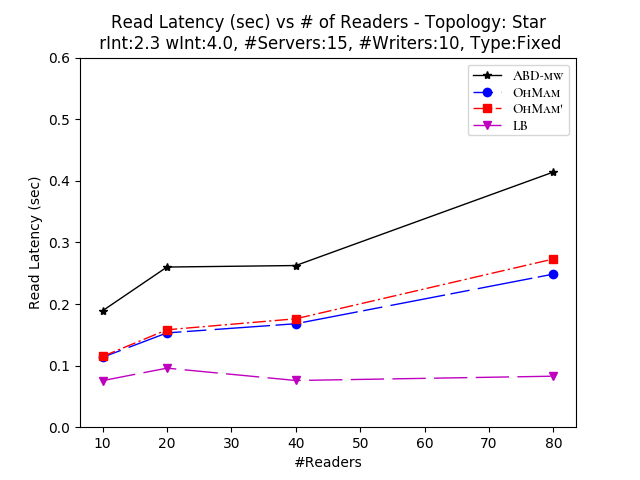} 
		&		 
			\includegraphics[width=0.5\textwidth]{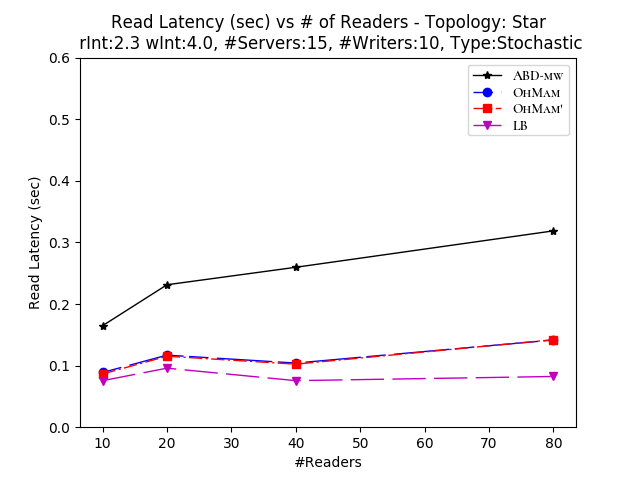}\\
				(g) & (h) \\ 
\end{tabular}
}
\vspace{-\smallskipamount}
\caption{{\small MWMR} Simulation Results.}
\label{fig:oram:mwmrplots}
\vspace{-\smallskipamount}
\end{figure}

\textit{Contention:} 
We examine the efficiency of our algorithms under different
concurrency schemes.
%
We set the operation frequency to be constant and we observe that in the \emph{stochastic} 
scheme  read operations complete faster 
than in the \emph{fixed} scheme;
see Figures \ref{fig:oram:swmrplots}(c) and  \ref{fig:oram:swmrplots}(d) for the {\small SWMR} setting, and 
Figures \ref{fig:oram:mwmrplots}(g) and  \ref{fig:oram:mwmrplots}(h) for the {\small MWMR} setting.
This is expected 
as the \emph{fixed} scheme causes congestion.
For the \emph{stochastic} scheme the invocation time intervals are
distributed uniformly, this reduces congestion and improves latency. 

\textit{Topology:} 
Figures \ref{fig:oram:swmrplots}(a) and \ref{fig:oram:swmrplots}(b) for the  {\small SWMR} setting, and 
Figures \ref{fig:oram:mwmrplots}(e) and \ref{fig:oram:mwmrplots}(f) for the {\small MWMR} setting 
show that topology substantially impacts  performance. 
For both the  {\small SWMR} and {\small MWMR} settings our algorithms outperform
algorithms \ABD{} and \ABDmwmr{} by a factor of at least 2 
in  \emph{Star} topology
where servers are well-connected.
Our   {\small SWMR} algorithms perform much better than \ABD{} also
in the \emph{Series} topology. 
For the {\small MWMR} setting and \emph{Series} topology, we note that \ABDmwmr{}
generally outperforms algorithm \omam{}, however the revised algorithm \omamex{} 
noticeably outperforms
\ABDmwmr{}.

Lastly we compare the performance of algorithms  \osam{} and \omam{} with
revised versions \osamex{} and \omamex{}.
We note that  \osamex{} and \omamex{} outperform 
all other algorithms in \emph{Series} topologies.
However, and perhaps not surprisingly, \osam{} and \omam{} outperform \osamex{} and \omamex{} 
in \emph{Star} topology. 
This is explained as follows.
In \emph{Star} topology  \emph{readRelay}
and  \emph{readAck} messages are exchanged quickly at the servers and thus 
delivered quickly to the clients.
On the other hand, the bookkeeping mechanism used in the revised algorithms 
incur additional computational latency, resulting in worse latency. 

An important observation is that while algorithms  \osamex{} and \omamex{}
improve the latencies of some operations (allowing some reads
to complete in two exchanges), their  performance relative to
algorithms   \osam{} and \omam{} depends on the deployment setting.
Simulations show that  \osam{} and \omam{}
are more suitable for datacenter-like deployment, while
in the ``looser''  settings 
 algorithms \osamex{} and \omamex{} perform better.

%

\section{Conclusions}
\label{sec:oram:conclude}

We focused on the problem of emulating atomic read/write shared objects 
in message-passing settings with the goal of using three communication exchanges
(to the extent allowed by the impossibility result \cite{HNS2017}).
We presented algorithms for the {\small SWMR} and {\small MWMR} models.
We then revised the algorithms to speed up some read operations.
We rigorously reasoned about the correctness of our algorithms. 
The algorithms do not impose any constrains on the number of readers 
({\small SWMR} and {\small MWMR}) and on the number of
the writers for the {\small MWMR} model. 
Finally we performed an empirical study of the performance
of algorithms using simulations.

\bibliographystyle{acm}
\bibliography{biblio}

\end{document}